\documentclass[10pt]{article}
\usepackage{amsmath, amssymb, slashed, latexsym, cite, mathrsfs, tensor, tikz-feynman, dsfont, simplewick, xcolor, faktor, float, comment, bbm}
\usepackage[a4paper, left=2.5cm, right=2.5cm, top=3.5cm, bottom=3cm]{geometry}
\usepackage[hidelinks]{hyperref}
\hypersetup{
	colorlinks=true,
	linkcolor=blue,
	citecolor=magenta,
}
\usepackage{booktabs,dcolumn,caption}
\numberwithin{equation}{section}
\makeatletter
\DeclareFontFamily{OMX}{MnSymbolE}{}
\DeclareSymbolFont{MnLargeSymbols}{OMX}{MnSymbolE}{m}{n}
\SetSymbolFont{MnLargeSymbols}{bold}{OMX}{MnSymbolE}{b}{n}
\DeclareFontShape{OMX}{MnSymbolE}{m}{n}{
	<-6>  MnSymbolE5
	<6-7>  MnSymbolE6
	<7-8>  MnSymbolE7
	<8-9>  MnSymbolE8
	<9-10> MnSymbolE9
	<10-12> MnSymbolE10
	<12->   MnSymbolE12
}{}
\DeclareFontShape{OMX}{MnSymbolE}{b}{n}{
	<-6>  MnSymbolE-Bold5
	<6-7>  MnSymbolE-Bold6
	<7-8>  MnSymbolE-Bold7
	<8-9>  MnSymbolE-Bold8
	<9-10> MnSymbolE-Bold9
	<10-12> MnSymbolE-Bold10
	<12->   MnSymbolE-Bold12
}{}

\let\llangle\@undefined
\let\rrangle\@undefined
\DeclareMathDelimiter{\llangle}{\mathopen}%
{MnLargeSymbols}{'164}{MnLargeSymbols}{'164}
\DeclareMathDelimiter{\rrangle}{\mathclose}%
{MnLargeSymbols}{'171}{MnLargeSymbols}{'171}
\makeatother

\newcommand{\im}{\mathrm{i}}
\newcommand{\ep}{\mathrm{e}}
\newcommand{\diff}{\mathrm{d}}
\newcommand{\tr}{\mathrm{tr}}
\newcommand{\sfrac}[2]{{\textstyle\frac{#1}{#2}}}
\newcommand{\+}{\dagger}
\renewcommand{\=}{\ =\ }
\newcommand{\unity}{\mathbbm{1}}
\newcommand{\und}{\quad\textrm{and}\quad}
\newcommand{\with}{\quad\textrm{with}\quad}

\newcommand{\A}{\widetilde A}
\newcommand{\R}{\widetilde R}
\newcommand{\G}{\mathcal{G}}
\newcommand{\rR}{\mathrm{R}}

\newcommand{\drm}{\ensuremath{\mathrm{d}}}
\newcommand{\blambda}{\ensuremath{\bar{\lambda}}}
\newcommand{\intdx}{\ensuremath{\int \drm^4 x\;}}
\newcommand{\bC}{\ensuremath{\bar{C}}}

\newcommand{\partialA}{\ensuremath{\partial\cdot A}}
\newcommand{\FP}{\ensuremath{\Delta_{\mathrm{FP}}[A]}}
\newcommand{\MSS}{\ensuremath{\Delta_{\mathrm{MSS}}[A]}}

\newcommand{\AT}{\ensuremath{A^{\mathrm{T}}}}
\newcommand{\AL}{\ensuremath{A^{\mathrm{L}}}}

\newcommand{\tA}{\ensuremath{\widetilde{A}}}

\newcommand{\D}{\mathcal{D}}
\newcommand{\tD}{\ensuremath{\widetilde{\mathcal{D}}}}
\newcommand{\tP}{\ensuremath{\widetilde{P}}}
\newcommand{\trD}{\ensuremath{\widetilde{\rD}}}
\newcommand{\tC}{\ensuremath{\widetilde{C}}}
\newcommand{\tbC}{\ensuremath{\widetilde{\bar{C}}}}
\newcommand{\tG}{\ensuremath{\widetilde{G}}}
\newcommand{\tS}{\ensuremath{\widetilde{S}}}
\newcommand{\tF}{\ensuremath{\widetilde{F}}}
\newcommand{\tFaux}{\ensuremath{\widetilde{F}}}
\newcommand{\tvarphi}{\ensuremath{\widetilde{\varphi}}}

\newcommand{\tlambda}{\widetilde{\lambda}}
\newcommand{\tblambda}{\widetilde{\bar{\lambda}}}
\newcommand{\tR}{\widetilde{R}}

\newcommand{\tPhi}{\widetilde{\Phi}}

\newcommand{\dPhi}{\Phi^{\+}}
\newcommand{\tpsi}{\widetilde{\psi}}
\newcommand{\tscrA}{\tilde{\mathscr{A}}}
\newcommand{\scrA}{\mathscr{A}}
\newcommand{\tscrD}{\widetilde{\mathscr{D}}}
\newcommand{\sscrD}{\slashed{\mathscr{D}}}
\newcommand{\stscrD}{\widetilde{\slashed{\mathscr{D}}}}
\newcommand{\scrD}{\mathscr{D}}
\newcommand{\scrC}{\mathscr{C}}
\newcommand{\scrCb}{\bar{\mathscr{C}}}
\newcommand{\hscrC}{\widehat{\mathscr{C}}}
\newcommand{\tRl}{\ensuremath{\stackrel{\leftarrow}{\tR}}}

\newcommand{\slA}{\slashed{A}}
\newcommand{\sltA}{\widetilde{\slashed{A}}}
\newcommand{\slAL}{\slashed{\AL}}

\newcommand{\nc}{n_{\mathrm{c}}}
\newcommand{\rD}{\mathrm{D}}
\renewcommand{\t}{\times}
\newcommand{\Diff}{\mathrm{D}}
\newcommand{\tDiff}{\ensuremath{\widetilde{\rD}}}
\newcommand{\chP}{\mathrm{P}}
\newcommand{\bpsi}{{\bar{\psi}}}
\newcommand{\tbpsi}{\widetilde{{\bar{\psi}}}}
\newcommand{\bDiff}{\bar{\mathrm{D}}}
\newcommand{\bW}{\bar{W}}
\newcommand{\dalpha}{{\dot{\alpha}}}
\newcommand{\dbeta}{{\dot{\beta}}}
\newcommand{\thet}{\theta}
\newcommand{\bthet}{{\bar{\theta}}}
\newcommand{\bsigma}{{\bar{\sigma}}}
\newcommand{\Lag}{\mathcal{L}}
\newcommand{\balpha}{\bar{\alpha}}
\newcommand{\bxi}{\bar{\xi}}
\newcommand{\bchi}{\bar{\chi}}
\newcommand{\Dc}{\ensuremath{\mathring{\Delta}}}

\newcommand{\Sten}{S^{\mathrm{(10)}}}
\newcommand{\Sfour}{S^{\mathrm{(4)}}}
\newcommand{\SU}{\mathrm{SU}}
\newcommand{\diag}{\operatorname{diag}}
\newcommand{\sdiagram}[5]{
	\begin{tikzpicture}[transform shape, scale=0.55, every node/.style={scale=1.3}, baseline={([yshift=-.1cm]current bounding box.center)}]
		\begin{feynman}
			\vertex (a);
			\vertex [right=of a] (b);
			\vertex [above left=of a] (d) {\(#1\)};
			\vertex [below left=of a] (e) {\(#2\)};
			\vertex [above right=of b] (f) {\(#4\)};
			\vertex [below right=of b] (g) {\(#5\)};
			\diagram* {
				(a) -- [edge label=\(#3\)] (b),
				(a) -- [photon] (d),
				(a) -- [photon] (e),
				(b) -- [photon] (f),
				(b) -- [photon] (g),
			};
		\end{feynman}\;
	\end{tikzpicture}
}
\newcommand{\xdiagram}[4]{
	\begin{tikzpicture}[transform shape, scale=0.5, every node/.style={scale=1.5}, baseline={([yshift=-.1cm]current bounding box.center)}]
		\begin{feynman}
			\vertex (a);
			\vertex [above left=of a] (d) {\(#1\)};
			\vertex [below left=of a] (e) {\(#2\)};
			\vertex [above right=of a] (f) {\(#3\)};
			\vertex [below right=of a] (g) {\(#4\)};
			\diagram* {
				(a) -- [photon] (d),
				(a) -- [photon] (e),
				(a) -- [photon] (f),
				(a) -- [photon] (g),
			};
		\end{feynman}\;
	\end{tikzpicture}
}
\newcommand*\circled[1]{\tikz[baseline=(char.base)]{
		\node[shape=circle,draw,inner sep=2pt] (char) {#1};}}

\begin{document}
	\sloppy
	\title{\bf\huge The coupling flow of\\[6pt]
		${\cal N}=\,4$ super Yang--Mills theory}
	\date{~}
	
	\author{\phantom{.}\\[12pt]
		{\scshape\Large Maximilian Rupprecht\footnote{\href{mailto:maximilian.rupprecht@itp.uni-hannover.de}{maximilian.rupprecht@itp.uni-hannover.de}}}
		\\[24pt]
		Institut f\"ur Theoretische Physik\\ 
		and\\ Riemann Center for Geometry and Physics
		\\[8pt]
		Leibniz Universit\"at Hannover \\ 
		Appelstra{\ss}e 2, 30167 Hannover, Germany
		\\[24pt]
	} 
	
	\clearpage
	\maketitle
	\thispagestyle{empty}
	
	\begin{abstract}
		\noindent\large
		We offer a novel perspective on ${\cal N}=\,4$ supersymmetric Yang--Mills (SYM) theory through the framework of the Nicolai map, a transformation of the bosonic fields that allows one to compute quantum correlators in terms of a free, purely bosonic functional measure. Generally, any Nicolai map is obtained through a path-ordered exponential of the so-called coupling flow operator.
		The latter can be canonically constructed in any gauge using an ${\cal N}=\,1$ off-shell superfield formulation of ${\cal N}=\,4$ SYM, or alternatively through dimensional reduction of the result from ${\cal N}=\,1$ $D=10$ SYM, in which case we need to restrict to the Landau gauge. We propose a general theory of the ${\cal N}=\,4$ coupling flow operator, arguing that it exhibits an ambiguity in form of an R-symmetry freedom given by the Lie algebra $\mathfrak{su}(4)$. This theory incorporates our two construction approaches as special points in $\mathfrak{su}(4)$ and defines a broad class of Nicolai maps for ${\cal N}=\,4$ SYM.
	\end{abstract}
	\newpage
	\tableofcontents
	\newpage
	\section{Introduction}
	\subsection{History and overview}
	Gauge theories are our best candidates for describing Nature at elementary scales. Out of all such theories, maximally supersymmetric Yang--Mills theory in four dimensions ($\mathcal{N}=\,4$ SYM) takes on a special role. In a sense it is the most simple (possibly integrable) gauge theory that can be formulated. Indeed it has been a long standing goal of theoretical physicists to solve this particular theory analytically. While being a toy model, it is expected that progress on this matter would have a drastic impact on our understanding of more complex models of nature. It is therefore important to continue this pursuit and investigate $\mathcal{N}=\,4$ SYM from every possible angle.
	
	A so far relatively unknown approach to supersymmetric field theories is referred to as the Nicolai map. It is based on works from Nicolai, Dietz, Flume and Lechtenfeld \cite{Nic1, Nic2, Nic3, L1, FL, DL1, DL2} from the 1980s. The Nicolai map is a transformation of the bosonic Yang--Mills fields that relates the interacting theory at some coupling $g$ to the free theory at zero coupling. It allows one to compute quantum correlators in terms of a free, purely bosonic functional measure, entirely bypassing the use of any anticommuting (Grassmann) variables. After a pause of roughly 35 years there has recently been renewed interest on this matter starting with the papers \cite{ANPP, NP, ALMNPP, AMPP}. Shortly after, a general formula for the Nicolai map in terms of the so-called coupling flow operator $R_g$ was found in \cite{LR1}. For theories that have an off-shell superfield formalism the coupling flow operator can be constructed canonically. In these cases, a general construction method for the Nicolai map in arbitrary gauges was developed in \cite{LR2} and independently in \cite{MN}. This framework has potential applications in all kinds of supersymmetric theories, e.g.~as recently investigated \cite{LN} in supermembrane and matrix theory.
	
	We should highlight that the Nicolai map and its coupling flow operator are (depending on the theory) not necessarily unique. The original proof \cite{Nic1,Nic2} by Nicolai only shows that there exists such a map. The non-uniqueness was seen most strikingly in \cite{AMPP}, where it was shown that there exist two distinct Nicolai maps in ${\cal N}=\,1$ $D=6$ SYM (at least to third order in the coupling). In fact, one of the main results of this work is that in case of ${\cal N}=4$ SYM, there is a 15-dimensional ambiguity in the coupling flow operator. However, by definition of the map, correlators are independent of the particular choice of the Nicolai map (or coupling flow operator).
	
	To this date, most works on Nicolai maps were restricted to $\mathcal{N}=\,1$ supersymmetry. Only in \cite{NP} a Nicolai map for $\mathcal{N}=\,4$ SYM was deduced by dimensional reduction from the map for $\mathcal{N}=\,1$ $D=10$ SYM. In our work, we develop a more extensive framework for dealing with $\mathcal{N}=\,4$ supersymmetry\footnote{other theories with extended supersymmetry such as $\mathcal{N}=\,2$ $D=6$ SYM could be described in an analogous fashion.}. Generally, we distinguish two possibilities for obtaining Nicolai maps in $\mathcal{N}=\,4$ SYM. The first one is dimensional reduction from ten to four dimensions, while the second one makes use of an $\mathcal{N}=\,1$ off-shell superfield formalism for $\mathcal{N}=\,4$ SYM. The former approach treats all four supersymmetries on an equal footing, whereas the latter singles out one of the supersymmetries. This leads to an ambiguity in the coupling flow operator and the corresponding Nicolai map. We explain this by an analysis of the R-symmetry of $R_g$. Essentially, the operator is subject to $\SU(4)$ R-symmetry transformations and to a principle of superpositions with weight one. This suggests a general description of the coupling flow operator in terms of an $\mathfrak{su}(4)$ R-symmetry freedom that incorporates the two before-mentioned results as special points in the Lie algebra.
	
	In Section \ref{sec:action} we compare two formulations of the $\mathcal{N}=\,4$ SYM action and how they are actually equivalent. We start with the $\mathcal{N}=1$ superfield formalism and then consider dimensional reduction from $\mathcal{N}=\,1$ $D=10$ SYM to $\mathcal{N}=\,4$ $D=4$ SYM. In Section \ref{sec:roperator} we compare the two corresponding formulations of the coupling flow operator and place them as special cases within a general R-symmetric understanding of the operator. In Section \ref{sec:nmaps} we investigate the resulting Nicolai maps. Finally, in Section \ref{sec:conclusions} we give our conclusions and an outlook to possible future directions that this work may point at.
	
	Due to the technical nature of this paper, we present many of the calculations in detailed appendices \ref{app:superfields}-\ref{app:tests} in order not to disrupt the common theme of our arguments.
	
	\subsection{Basics of the Nicolai map}
	To begin with, we recall the essentials of the Nicolai map, without specializing to a particular theory. Any supersymmetric theory can be expressed in terms of bosonic and fermionic (potentially including ghost) fields. Usually the latter appear quadratically so that they can be integrated out, giving a nonlocal functional determinant. The resulting action can be written as
	\begin{equation}\label{eq:nonl_action}
		S_g[\phi]\=S_g^{\mathrm{b}}[\phi]+\hbar\  S_g^{\mathrm{f}}[\phi]\ ,
	\end{equation}
	with coupling constant $g$ and local, nonlocal parts of the action $S_g^{\mathrm{b}}$, $S_g^{\mathrm{f}}$ respectively. Here, $\phi$ stands for the bosonic field content of the theory, i.e.~for $\mathcal{N}=\,4$ SYM we have $\phi=(A_\mu,\ \varphi_i)$ with the gauge field $A_\mu$ and six real scalars $\varphi_i$. Expectation values\footnote{by the vanishing of the vacuum energy in supersymmetric theories, we have the normalization $\langle 1\rangle_g=1$.} of bosonic observables $X[\phi]$ in the theory \eqref{eq:nonl_action} are given by
	\begin{equation}
		\langle X[\phi] \rangle_g\=\int \mathcal{D}\phi\ \exp\bigl\{\sfrac{\im}{\hbar}S_g[\phi]\bigr\}\ X[\phi]\ .
	\end{equation}
	The Nicolai map is a (nonlinear and nonlocal) field transformation
	\begin{equation}
		T_g:\ \phi(x)\ \mapsto\ \phi'(x;\ g, \phi)\ ,
	\end{equation}
	invertible at least as a formal power series in $g$, with the defining property
	\begin{equation}\label{eq:def_nmap}
		\langle X[\phi] \rangle_g\=\langle X[T_g^{-1}\phi]\rangle_0\qquad\forall X\ ,
	\end{equation}
	that connects the interacting theory at coupling $g$ with the free theory ($g=0$). We stress again that the map $T_g$ is not necessarily unique. However, since we construct all maps from the defining relation \eqref{eq:def_nmap}, correlators do not depend on the choice of the particular map.
	Taking the derivative of \eqref{eq:def_nmap} with respect to the coupling gives
	\begin{equation}\label{eq:def_cflow}
		\partial_g \langle X[\phi] \rangle_g\=\langle (\partial_g+R_g[\phi])X[\phi] \rangle_g\ ,
	\end{equation}
	which defines the infinitesimal version of the Nicolai map, the so-called coupling flow operator
	\begin{equation}
		R_g[\phi] \= \int\!\diff x\ \bigl(\partial_g T_g^{-1} \circ T_g \bigr) \phi(x)\,\frac{\delta}{\delta\phi(x)}\ =:\ \int\!\diff x\ K[\phi;\;x]\,\frac{\delta}{\delta\phi(x)}\ ,
	\end{equation}
	with kernel $K$. By setting $X[\phi]=T_g\phi$ in \eqref{eq:def_nmap}, one can quickly derive \cite{L2} the relation
	\begin{equation}
		(\partial_g+R_g[\phi])T_g\phi\=0\ .
	\end{equation}
	This is a well-known differential equation solved by the path-ordered exponential
	\begin{equation} \label{eq:closedT}
		T_g\,\phi \= \overrightarrow{\cal P} \exp \Bigl\{-\!\int_0^g\!\diff h\ R_h[\phi]\Bigr\}\ \phi\ ,
	\end{equation}
	which was first found in \cite{LR1}. This shows that the knowledge of $R_g$ completely captures the analytic $g$-dependence of the Nicolai map, allowing its perturbative construction. It is the main objective of this work to find the explicit and most general form of $R_g$ for $\mathcal{N}=\,4$ SYM. 
	
	Next, we note the characteristic properties\footnote{which were originally used as the defining conditions of the Nicolai map, but can be traded for the single relation \eqref{eq:def_nmap}.} of the Nicolai map and the corresponding infinitesimal properties for the coupling flow operator. Writing \eqref{eq:def_nmap} in terms of path integrals and collecting powers of $\hbar$, one finds
	\begin{equation}
		S^{\mathrm{b}}_0[T_g\phi] \= S^{\mathrm{b}}_g[\phi] \quad\und\quad
		S^{\mathrm{f}}_0[T_g\phi] -\im\,\tr\ln\sfrac{\delta T_g\phi}{\delta\phi} \= S^{\mathrm{f}}_g[\phi]\ ,
	\end{equation}
	the `free-action' and `determinant-matching' conditions respectively. For gauge theories, we have the additional property that the chosen gauge fixing function $\G(\phi)$ is a fixed point of the Nicolai map. From \eqref{eq:def_cflow}, it is straightforward to deduce \cite{L2} the corresponding infinitesimal conditions
	\begin{equation}\label{eq:cf_cond12}
		(\partial_g+R_g[\phi])S^{\mathrm{b}}_g[\phi]\=0 \quad\und\quad (\partial_g+R_g[\phi])S^{\mathrm{f}}_g[\phi]\=\int\!\diff x\ \frac{\delta K[\phi;\;x]}{\delta \phi(x)}\ ,
	\end{equation}
	as well as the gauge condition
	\begin{equation}\label{eq:cf_cond3}
		(\partial_g+R_g[\phi])\G(\phi)\=0\ .
	\end{equation}

	For completeness, although it will not be relevant to the rest of this work, we include here a few general remarks on regularization and renormalization. Since the Nicolai map itself only consists of tree graphs, regularization is not required at this stage. Only in the end, when computing correlators in the free theory with \eqref{eq:def_nmap}, one has to contract trees with each other. This generates loops (but interestingly, none of them purely fermionic) that have to be regularized. In the case of ${\cal N}=\,1$ SYM this technique and the subsequent renormalization are successfully carried out in the paper \cite{DL2} from 1985, for example rederiving the universality of the gauge coupling to 1-loop order. The computational effort of this method as opposed to the traditional Feynman diagram approach is practically comparable.
	
	\subsection{Conventions and notation}
	In this paper, we work in four- and sometimes ten-dimensional Minkowski space equipped with the mostly plus metric
	\begin{equation}
		\eta^{\mu\nu}\=\diag(-1,\ +1,\ +1,\ +1)\ ,\qquad \eta^{\Sigma\Theta}\=\diag(-1,\ +1,\ ...\ ,\ +1)
	\end{equation}
	respectively, where lowercase Greek indices run from 0 to 3 and uppercase Greek indices run from 0 to 9. For the four-dimensional spinor algebra, we adopt the conventions from \cite{WB}, including the chiral basis for the gamma matrices
	\begin{equation}
		\gamma^\mu\=\begin{pmatrix}0 & \sigma^\mu \\ \bsigma^\mu & 0\end{pmatrix}\ ,\qquad \gamma^5\=\gamma^0\gamma^1\gamma^2\gamma^3\=\begin{pmatrix}-\im & 0 \\ 0 & \im\end{pmatrix}\ ,
	\end{equation}
	with sigma matrices
	\begin{equation}
		\sigma^0\=\begin{pmatrix}-1 & 0 \\ 0 & -1\end{pmatrix}\ ,\qquad \sigma^1\=\begin{pmatrix}0 & 1 \\ 1 & 0\end{pmatrix}
		\ ,\qquad \sigma^2\=\begin{pmatrix}0 & -\im \\ \im & 0\end{pmatrix} ,\qquad \sigma^3\=\begin{pmatrix}1 & 0 \\ 0 & -1\end{pmatrix}\ ,
	\end{equation}
	and $\bsigma^0=\sigma^0$, $\bsigma^{1,2,3}=-\sigma^{1,2,3}$. The gamma matrices satisfy the Clifford algebra $\{\gamma^\mu,\ \gamma^\nu\}=-2\eta^{\mu\nu}$.
	In this basis, the chiral projectors $\chP^\pm$ take the form
	\begin{equation}
		\chP^+\=\sfrac12 (1+ \im\gamma^5)\=\begin{pmatrix}1 & 0 \\ 0 & 0\end{pmatrix}\ ,\qquad \chP^-\=\sfrac12 (1 - \im\gamma^5)\=\begin{pmatrix}0 & 0 \\ 0 & 1\end{pmatrix}\ .
	\end{equation}
	We often use the standard Feynman slash notation
	\begin{equation}
		\slashed{a}\=\gamma^\mu a_\mu\ ,
	\end{equation}
	with the exception that the slashed script letters $\slashed{\scrA}$ and $\sscrD$ have a related but distinct meaning that is defined in the main text. We additionally define the antisymmetric
	\begin{equation}
		\gamma^{\mu\nu}\=\sfrac{1}{2}(\gamma^{\mu}\gamma^{\nu}-\gamma^{\nu}\gamma^{\mu})
	\end{equation}
	and generally antisymmetrize indices with weight one, indicated by square brackets, e.g.~
	\begin{equation}
		a^{[\mu}b^{\nu]}\=\sfrac12 (a^\mu b^ \nu-a^\nu b^\mu)\ .
	\end{equation}
	All of our fields are in the adjoint representation of the gauge group which we take to be $\SU(\nc)$ with real antisymmetric structure constants
	$f^{abc}$ such that
	\begin{equation}
		f^{abc}f^{abd}\=\nc\delta^{cd}\ ,
	\end{equation}
	where color indices run from 1 to $\nc^2-1$, which we often leave implicit. For example, we write the non-abelian field-strength tensor $F_{\mu\nu}$ in two equivalent notations
	\begin{equation}\label{eq:field_strength}
		F_{\mu\nu}\=\partial_\mu A_\nu-\partial_\nu A_\mu+gA_\mu\times A_\nu\qquad\iff\qquad F^a_{\mu\nu}\=\partial_\mu A^a_\nu-\partial_\nu A^a_\mu+gf^{abc}A^b_\mu A^c_\nu\ ,
	\end{equation}
	and the covariant derivative
	\begin{equation}\label{eq:cov_der}
		\Diff_\mu\=\partial_\mu+gA_\mu\times\ \qquad\iff\qquad (\Diff_\mu...)^a\=\partial_\mu(...)^a+gf^{abc}A^b_\mu(...)^c\ .
	\end{equation}
	We sum over implicit color indices of products, e.g.
	\begin{equation}
		F_{\mu\nu}F^{\mu\nu}\=F^a_{\mu\nu}F^{a\,\mu\nu}\ ,
	\end{equation}
	except when we write an explicit cross product, e.g.
	\begin{equation}
		\varphi\ \psi {\t} \lambda \= f^{abc}\varphi^a \psi^b \lambda^c\ .
	\end{equation}
	Further, e.g.~when writing down Nicolai maps to second order, we often adopt from section 4 of \cite{ALMNPP} the shorthand notations for multiplying quantities in color and position space. This means that all objects are multiplied as color matrices or vectors, and integration kernels are convoluted with insertions of bosonic fields $A_\mu$ or $\varphi_i$ for $\mu=0,1,2,3$ and $i=1,...,6$. For example, we would write in two equivalent notations the expression
	\begin{equation}
		\partial^\rho C \varphi_i \partial_\mu C A_\rho{\t} \varphi_i \qquad\iff\qquad \int \drm^4y\;\drm^4z\   \partial^\rho C(x-y)\; (f^{abc}\varphi^b_i)(y)\; \partial_\mu C(y-z)\; (f^{cde}A^d_\rho)(z) \varphi^e_i(z)\ ,
	\end{equation}
	with the scalar propagator $C=\Box^{-1}$. We often summarize the bosonic fields in the symbol
	\begin{equation}
		\scrA_\Gamma\=(A_\mu,\ \varphi_i)\ .
	\end{equation}
	An overview over the various types of indices to be used in the following can be found in Table \ref{tab:indices}.
	
	\begin{table}[h]
		\setlength\tabcolsep{0pt}
		\caption{Types of indices used in this paper. Color and spinor indices are often left implicit.}
		\label{tab:indices}
		\begin{tabular*}{\textwidth}{@{\extracolsep{\fill}} lccc }
			\toprule
			\multicolumn{1}{l}{Name} & \multicolumn{1}{c}{Representation} & \multicolumn{1}{c}{Range} & \multicolumn{1}{c}{Alphabet}\\
			\midrule
			R-symmetry & $\mathbf{4}$ of $\SU(4)$ & 1 to 4 & 1st half of uppercase Latin ($A,B,C,...$)\\
			R-symmetry (broken) & $\mathbf{3}$ of $\SU(3)$ & 1 to 3 & 2nd half of uppercase Latin ($I,J,K,...$)\\
			R-symmetry & $\mathbf{6}$ of $\SU(4)\ \cong\ \operatorname{SO}(6)$ & 1 to 6 & 2nd half of lowercase Latin ($i,j,k,...$)\\
			Color & Adjoint of $\SU(\nc)$ & 1 to $\nc^2-1$ & 1st half of lowercase Latin ($a,b,c,...$)\\
			Lorentz (4-dim.) & Spin 1 of $\operatorname{SO}(1,3)$ & 0 to 3 & 2nd half of lowercase Greek ($\mu, \nu, \rho, ...$)\\
			Lorentz (10-dim.) & Spin 1 of $\operatorname{SO}(1,9)$ & 0 to 9 & uppercase Greek ($\Sigma, \Theta, \Gamma, ...$)\\
			Spinor & Spin $\sfrac12$ & 1 to 4 & 1st half of lowercase Greek ($\alpha, \beta, \gamma, ...$)\\
			\bottomrule
		\end{tabular*}
	\end{table}
	
	Lastly, we recall the definitions of the basic building blocks of the coupling flow operator in $\mathcal{N}=\,1$ $D=4$ SYM \cite{LR2}, since these also appear in the more complicated $\mathcal{N}=\,4$ case. The free gaugino and ghost propagators $S_0$ and $G_0$ which will be used for the perturbative expansion of their full versions are given by 
	\begin{equation}
		S_0\=\slashed{\partial}^{\;-1}\=-\slashed{\partial}C\ ,\qquad G_0\=(\sfrac{\partial \G(\scrA)}{\partial A_\mu}\partial_\mu)^{-1}\ ,
	\end{equation}
	respectively with the gauge fixing function $\G(\scrA)$. 
	An object that appears often in the context of the Nicolai map is the free projector
	\begin{equation}
		\Pi{\indices{_\mu^\nu}}\=\delta\indices{_\mu^\nu}-\partial_\mu G_0\sfrac{\partial\G(\scrA)}{\partial \scrA_\nu}\ ,
	\end{equation}
	which we often extend to capital greek indices 
	\begin{equation}
		\Pi{\indices{_\Gamma^\Sigma}}\=\delta\indices{_\Gamma^\Sigma}-\partial_\Gamma G_0\sfrac{\partial\G(\scrA)}{\partial \scrA_\Sigma}\ ,
	\end{equation}
	with the understanding that in the reduced four-dimensional theory $\partial_{3+i}=0$ for $i=1,...,6$. Among all possible gauges, the Landau gauge $\G(\scrA)=\partial^\mu A_\mu$ takes on a special role, since in that case $G_0\equiv C$ so that $\Pi\indices{_\mu^\nu}$ equals the standard transversal projector
	\begin{equation}
		\amalg\indices{_\mu^\nu}\=\delta\indices{_\mu^\nu}-\partial_\mu C\partial^\nu\ .
	\end{equation} 
	The latter splits the Yang--Mills fields into transversal and longitudinal components
	\begin{equation}\label{eq:split_T_L}
		A_\mu\=\AT_\mu+\AL_\mu\ ,\qquad \AT_\mu\=\amalg\indices{_\mu^\nu}A_\nu\ ,\qquad \AL_\mu\=(\delta\indices{_\mu^\nu}-\amalg\indices{_\mu^\nu})A_\nu\=\partial_\mu C\ \partial\cdot A\ ,
	\end{equation}
	where we abbreviate $\partialA=\partial^\mu A_\mu$. In arbitrary gauges (or outside of the gauge hypersurface of the Landau gauge), it is helpful to define the `conjugate' Yang--Mills field
	\begin{equation}\label{eq:conjugate_fields}
		A^*_\mu\ :=\ \AT_\mu-\AL_\mu\= A_\mu -2\partial_\mu C\partialA\ ,
	\end{equation}
	although for the most part in this work, we restrict ourselves to the Landau gauge hypersurface, where $A=A^*$.
	
	\section{$\mathcal{N}=\,4$ SYM action}\label{sec:action}
	A common formulation of the $\mathcal{N}=\,4$ SYM invariant action (without a topological term\footnote{which we neglect here for simplicity but could be included in a future analysis. It is expected to lead to an additional chiral freedom in our theory.}) is \cite{BSS}
	\begin{equation}\label{eq:action_Weyl}
		\begin{aligned}
			S_{\mathrm{inv}}\=\intdx \Bigl\{&-\sfrac14 F^{\mu\nu}F_{\mu\nu}-\sfrac12 \Diff_\mu \varphi_i\Diff^\mu\varphi_i-\sfrac{\im}{2}\bchi_A\slashed{\Diff}\chP^+\chi^A-\sfrac{\im}{2}\bar{\tilde{\chi}}^A\slashed{\Diff}\chP^-\tilde{\chi}_A\\
			&-\im g\ t\indices{^i_{AB}}\bar{\tilde{\chi}}^A\chP^+\varphi_i\t \chi^B
			+\im g\ t\indices{^{iAB}}\bar{\chi}_A\chP^-\varphi_i\t \tilde{\chi}_B
			-\sfrac{g^2}{4}(\varphi_i\t \varphi_j)^2\Bigr\}\ ,
		\end{aligned}
	\end{equation}
	in terms of Weyl spinors $\chi^A$, $\tilde{\chi}_A$ where $\tilde{\chi}_A=\mathrm{C}(\bar{\chi}^A)^{\mathrm{T}}$ with the charge conjugation operator $\mathrm{C}$ in four dimensions. All fields are in the adjoint representation of the gauge group, with color indices left implicit. Here, $\chi^A$ transforms as a $\mathbf{4}$ under the global $\SU(4)\ \cong\ \operatorname{SO}(6)$ R-symmetry, while $\tilde{\chi}_A$ transforms as a $\bar{\mathbf{4}}$. The indices $i=1,2,...,6$ label the six bosonic fields $\varphi_i$ that transform as a $\mathbf{6}$. Furthermore, the coefficients $t\indices{^i_{AB}}=(t\indices{^{iAB}})^*$ are the structure constants of the R-symmetry, or in other words Clebsch-Gordon coefficients that couple two $\mathbf{4}$'s to a $\mathbf{6}$ \cite{Kov1}. They allow us to define anti-symmetric complex scalars
	\begin{equation}\label{eq:Clebsch-Gordon}
		\varphi_{AB}\=t\indices{^i_{AB}}\varphi_i\ ,\qquad \varphi^{AB}\=t\indices{^{iAB}}\varphi_i\=(\varphi_{AB})^*
	\end{equation}
	and will be specified explicitly when we construct the action below.
	In this work, we find it advantageous to work with Majorana spinors instead of Weyl spinors. To that aim, we define
	\begin{equation}\label{eq:Majorana_red}
		\psi^A\=\chP^+\chi^A\ +\ \chP^-\tilde{\chi}_A\ ,\qquad \bpsi_A\=\bar{\chi}_A\chP^-\ +\ \bar{\tilde{\chi}}^A\chP^+\ .
	\end{equation}
	With $\mathrm{C}\gamma_5=\gamma_5 \mathrm{C}$, it is straightforward to check that $\psi^A=\mathrm{C}(\bar{\psi}_A)^{\mathrm{T}}$, which shows that $\psi^A$ are indeed Majorana spinors. A slight complication with this definition is that we need to be careful with R-symmetry transformations, since $\psi^A$ transforms neither as a $\mathbf{4}$ nor a $\bar{\mathbf{4}}$. We emphasize this point, because in this Majorana formulation, the position of the R-symmetry indices does not indicate the transformation properties of the corresponding quantities. When translating objects from the Weyl formulation to the Majorana formulation, index positions on the two sides of the equation do not match up. Hence, one generally has to remember the R transformation properties from the Weyl formulation. However, the Majorana formulation allows us to write the action in the more compact form
	\begin{equation}\label{eq:action_Maj}
			S_{\mathrm{inv}}\=\intdx \Bigl\{-\sfrac14 F^{\mu\nu}F_{\mu\nu}-\sfrac12 \Diff_\mu \varphi_i\Diff^\mu\varphi_i-\sfrac{\im}{2}\bpsi_A(\slashed{\Diff}\delta\indices{^A_B}+g\Phi\indices{^A_B}\t) \psi^B-\sfrac{g^2}{4}(\varphi_i\t \varphi_j)^2\Bigr\}\ ,
	\end{equation}
	with
	\begin{equation}\label{eq:Phi}
		\Phi\indices{^A_B}\ :=\ 2\bigl[t\indices{^i_{AB}}\chP^+-t\indices{^{iAB}}\chP^-\bigr]\varphi_i\ \equiv\ (c^i)\indices{^A_B}\varphi_i\ ,
	\end{equation}
	where we have defined a matrix-valued field $\Phi\indices{^A_B}$ that is obtained from the scalars $\varphi_i$ through contraction with matrix-valued coefficients $(c^i)\indices{^A_B}$.
	We further often use the shorthand
	\begin{equation}\label{eq:slscrD}
		\slashed{\scrD}{\indices{^A_B}}\ :=\ \slashed{\Diff}\delta\indices{^A_B}+g\Phi\indices{^A_B}\t\ .
	\end{equation}
	In the following two subsections, we show how the action \eqref{eq:action_Maj} is obtained from an $\mathcal{N}=\,1$ superfield formalism and from dimensional reduction, respectively.

	\subsection{$\mathcal{N}=\,1$ superfield formalism}
	It is well established that $\mathcal{N}=\,4$ SYM does not have a formulation in which all four supersymmetries are realized off-shell. However, it is possible to single out one of the supersymmetries to construct an $\mathcal{N}=\,4$ action using an $\mathcal{N}=\,1$ superfield formalism \cite{Kov1}. The field content resides in one vector superfield $V$ and three chiral superfields $\Phi_I$
	\begin{equation}
		V\=(A_\mu,\ \lambda,\ \D)\ ,\qquad \Phi_I\=(\phi_I,\ \psi_{I},\ F_I) \with I\=1,2,3\ .
	\end{equation}
	All fields are in the adjoint representation of the gauge group. The propagating degrees of freedom are the vector field $A_\mu$, four Weyl- (or equivalently Majorana-) spinors $\psi^A$ ($A=1,2,3,4$, with $\lambda=\psi^4$) and three complex scalars $\phi_I$. Further, there is one real scalar auxiliary field $\D$ and three complex scalar auxiliary fields $F_I$. In terms of superfields, the $\mathcal{N}=\,4$ Lagrangian density in Weyl notation is the last component\footnote{recall that $\thet^2\theta^\alpha=0$ and $\bthet^2\bthet_\dalpha=0$.} of a superfield:
	\begin{equation}\label{eq:Lag}
		\Lag\=\sfrac{1}{g^2\nc}\tr\Bigl[\sfrac{1}{16}\bigl(W^\alpha W_\alpha\bigr|_{\thet\thet}+\mathrm{h.c.}\bigr)\ +\ \ep^{-2V}\Phi^\+_I \ep^{2V} \Phi_I\bigr|_{\thet\thet\bthet\bthet}\ +\ \im \sfrac{\sqrt{2}}{3!}\bigl( \epsilon_{IJK}\Phi_I[\Phi_J,\Phi_K]\bigr|_{\thet\thet}+\mathrm{h.c.}\bigr)\Bigr]\ ,
	\end{equation}
	where $...|_{\theta\theta}$ denotes the $\theta\theta$-component of a given superfield and so on. The trace is over color space. We have also introduced the non-abelian supersymmetric field strength $W_\alpha$ and its conjugate
	\begin{equation}\label{eq:non-ab-susy-fs}
			W_\alpha\=-\sfrac{1}{4}\bDiff\bDiff \ep^{-2V}\Diff_\alpha \ep^{2V}\ ,\qquad
			\bW^\dalpha\=-\sfrac{1}{4}\Diff\Diff \ep^{-2V}\bDiff^\dalpha \ep^{2V}\ ,
	\end{equation}
	in chiral superspace, with the superspace covariant derivatives $\Diff_\alpha$, $\bDiff^\dalpha$.
	Note that the coupling only appears as an overall factor $1/g^2$ in front of \eqref{eq:Lag}.
	One recovers the usual dependence on the coupling by rescaling $V\ \rightarrow\ gV$ and $\Phi_I\ \rightarrow\ g\Phi_I$. Given the various superspace expansions, it is straightforward to obtain the Lagrangian explicitly in terms of components. The details on the computations that lead to the following results can be found in Appendix \ref{app:superfields}. We find for the Lagrangian in the Majorana basis
	\begin{equation}\label{eq:Lag_Maj}
		\begin{aligned}
			g^2 \Lag\=&-\sfrac{1}{4}F_{\mu\nu}F^{\mu\nu}-\sfrac{\im}{2}\blambda\gamma^\mu\Diff_\mu\lambda+\sfrac{1}{2}\D^2
			-\sfrac{1}{\sqrt{2}}\epsilon_{IJK}\bigl(F_I\phi_J{\t}\phi_K+F^{\+}_I\phi_J^{\+}{\t}\phi_K^{\+}\bigr)\\
			&-\Diff_\mu\phi^{\+}_I\Diff^\mu\phi_I-\sfrac{\im}{2}\bpsi_I\gamma^\mu\Diff_\mu\psi_I+F^{\+}_IF_I
			+\sfrac{1}{\sqrt{2}}\epsilon_{IJK}\bigl(\phi_I\bpsi_J\chP^+{\t}\psi_K+\phi^{\+}_I\bpsi_J\chP^-{\t}\psi_K\bigr)\\
			&-\sqrt{2}\bigl(\bpsi_I\chP^-\lambda{\t}\phi_I+\bpsi_I\chP^+\lambda{\t}\phi^{\+}_I\bigr)-\im \phi_I^{\+}\D{\t}\phi_I\ ,
		\end{aligned}
	\end{equation}
	and for the supersymmetry transformations
	\begin{equation}\label{eq:supervariations}
		\begin{aligned}
			&\delta_\alpha \phi_I\=\sqrt{2}(\bpsi_I\chP^+)_\alpha\ ,\\
			&\delta_\alpha \phi^{\+}_I\=\sqrt{2}(\bpsi_I\chP^-)_\alpha\ ,\\
			&\delta_\alpha (\chP^+\psi_I)_\beta \=-\im\sqrt{2}(\chP^+\gamma^\mu)_{\beta\alpha}(\Diff_\mu\phi_I)-\sqrt{2}(\chP^+)_{\beta\alpha} F_I\ ,\\
			&\delta_\alpha (\chP^-\psi_I)_\beta \=-\im\sqrt{2}(\chP^-\gamma^\mu)_{\beta\alpha}(\Diff_\mu\phi^{\+}_I)-\sqrt{2}(\chP^-)_{\beta\alpha} F^{\+}_I\ ,\\
			&\delta_\alpha F_I\=-\im\sqrt{2}(\Diff_\mu\bpsi_{I\beta}) (\gamma^\mu\chP^-)_{\beta\alpha}  - 2 \phi_I{\t}(\blambda\chP^-)_\alpha\ ,\\
			&\delta_\alpha F^{\+}_I\=-\im\sqrt{2}(\Diff_\mu\bpsi_{I\beta}) (\gamma^\mu\chP^+)_{\beta\alpha}  - 2 \phi_I^{\+}{\t}(\blambda\chP^+)_\alpha\ ,\\
			&\delta_\alpha A_\nu\=-\im(\blambda\gamma_\nu)_{\alpha}\ ,\\
			&\delta_\alpha\D\=-\im(\Diff_\mu \blambda_\beta)(\gamma_5 \gamma^\mu)_{\beta\alpha}\ ,\\
			&\delta_\alpha \lambda_\beta\=-\sfrac{1}{2}(\gamma^{\mu\nu})_{\beta\alpha}F_{\mu\nu}+\mathcal{D}(\gamma_5)_{\beta\alpha}\ .
		\end{aligned}
	\end{equation}
	The decisive advantage of the superfield formalism is that we can deduce the penultimate component\footnote{that is, the components with one less power of $\theta$ and $\bthet$ than maximal for the respective contributions.} of the superfield in \eqref{eq:Lag}. It reads
	\begin{equation}\label{eq:Dc_Maj}
		\begin{aligned}
			\Dc_\alpha\=&\sfrac14\intdx\Bigl\{-\D\gamma_5\lambda-\sfrac12 F_{\mu\nu}\gamma^{\mu\nu}\lambda+2\epsilon_{IJK}\bigl[\chP^+\psi_I\phi_J{\t}\phi_K+\chP^-\psi_I\phi^{\+}_J{\t}\phi^{\+}_K\bigr]+2\im \gamma_5\phi_I^{\+}\lambda{\t}\phi_I\\
			&+\im \sqrt{2}\bigl[\gamma^\mu\chP^-\psi_I\Diff_\mu\phi_I+\gamma^\mu\chP^+ \psi_I\Diff_\mu\phi^{\+}_I\bigr]-\sqrt{2}\bigl[\chP^+\psi_IF^{\+}_I+\chP^-\psi_IF_I\bigr]\Bigr\}_{\alpha}\ .
		\end{aligned}
	\end{equation}
	The superfield structure now enables us to write the invariant action as a supervariation
	\begin{equation}\label{eq:Sinv_3complex_off}
			S_{\mathrm{inv}}\=\intdx\ \Lag\=\sfrac{1}{2g^2}\ \delta_\alpha \Dc_\alpha\ ,
	\end{equation}
	which will be the central ingredient in the canonical construction of the coupling flow operator later on.
	To find the on-shell invariant action, we first need to insert the equations of motion for the auxiliary fields
	\begin{equation}\label{eq:aux}
		\mathcal{D}\=-\im \phi^{\+}_I{\t}\phi_I\ ,\qquad 
		F_I\=\sfrac{1}{\sqrt{2}}\epsilon_{IJK}\phi^{\+}_J{\t}\phi^{\+}_K\ ,
	\end{equation}
	resulting in
	\begin{equation}\label{eq:Sinv_3complex_on}
		\begin{aligned}
			S_{\mathrm{inv}}\=\sfrac{1}{g^2}\intdx\;  \Bigl\{-\sfrac{1}{4}F^{\mu\nu}F_{\mu\nu}-\Diff_\mu\phi^{\+}_I\Diff^\mu\phi_I-\sfrac{\im}{2}\bpsi_A\slashed{\Diff}\psi^A+\sfrac{1}{\sqrt{2}}\epsilon_{IJK}\bigl(\phi_I\bpsi_J\chP^+{\t}\psi_K+\phi^{\+}_I\bpsi_J\chP^-{\t}\psi_K\bigr)&\\
			-\sqrt{2}\bigl(\bpsi_I\chP^-\lambda{\t}\phi_I+\bpsi_I\chP^+\lambda{\t}\phi^{\+}_I\bigr)+\sfrac{1}{2}(\phi^{\+}_I\t\phi_I)^2-\sfrac{1}{2}\epsilon_{IJK}\epsilon_{ILM}(\phi_J\t\phi_K)(\phi^{\+}_L\t \phi^{\+}_M)\Bigr\}\ .&
		\end{aligned}
	\end{equation}
	Note that in this expression, the scalars are represented by three complex fields $\phi_I$. In order to get to the formulation \eqref{eq:action_Maj}, we need to replace these by six real fields $\varphi_i$ by a suitable identification
	\begin{equation}
		\phi_I\=\sfrac{1}{\sqrt2}(\varphi_{I+3}+\im\varphi_{I})\ ,\qquad \phi_I^{\+}\=\sfrac{1}{\sqrt2}(\varphi_{I+3}-\im \varphi_{I})\ ,
	\end{equation}
	giving
	\begin{equation}\label{eq:Sinv_6real_on}
		\begin{aligned}
			S_{\mathrm{inv}}\=\sfrac{1}{g^2}\intdx\;  \Bigl\{&-\sfrac{1}{4}F^{\mu\nu}F_{\mu\nu}-\sfrac12\Diff_\mu\varphi_i\Diff^\mu\varphi_i-\sfrac{\im}{2}\bpsi_A\slashed{\Diff}\psi^A\\
			+\sfrac12&\epsilon_{IJK}\bigl(\bpsi_I\varphi_{J+3}{\t}\psi_K-\bpsi_I\varphi_{J}\gamma_5{\t}\psi_K\bigr)+\bpsi_I\varphi_{I+3}{\t}\lambda+\bpsi_I\varphi_I\gamma_5{\t}\lambda-\sfrac{1}{4}(\varphi_i\t\varphi_j)^2\Bigr\}\ ,
		\end{aligned}
	\end{equation}
	where for the potential term the Jacobi identity in color space was used. From this expression we can read off the coefficients $(c^i)\indices{^A_B}$ from \eqref{eq:Phi}
	\begin{equation}\label{eq:cs}
		(c^I)\indices{^J_4}\=\im\delta_{IJ}\gamma_5\ ,\qquad (c^{I+3})\indices{^J_4}\=\im\delta_{IJ}\unity_4\ ,\qquad (c^I)\indices{^J_K}\=\im\epsilon_{IJK}\gamma_5\ ,\qquad(c^{I+3})\indices{^J_K}\=-\im\epsilon_{IJK}\unity_4\ ,
	\end{equation}
	wich are anti-symmetric under exchange of $A$ and $B$ and all others are zero. We will show in Section \ref{subsec:action_dimred} that these exactly match the coefficients obtained from dimensional reduction.
	
	To conclude this subsection, we note two results that will be used for the canonical construction of the coupling flow operator in Section \ref{subsec:cconstr_rop}. The susy-transformations of the six real scalars are
	\begin{equation}\label{eq:susy-trf-scalars}
		\delta_\alpha \varphi_i\=-\im \bpsi_J(c^i)\indices{^J_4}\ ,
	\end{equation}
	and the penultimate superfield component in terms of the six real scalars and with the auxiliary fields integrated out can be brought to the compact form
	\begin{equation}
		\Delta_\alpha\=\sfrac14\intdx\Bigl\{-\sfrac12 F_{\mu\nu}\gamma^{\mu\nu}\lambda-(\Phi\indices{^4_A})^{\+}\slashed{\scrD}\indices{^A_B}\psi^B+\sfrac12(\Phi\indices{^4_A})^{\+}\Phi\indices{^A_B}{\t}\psi^B\Bigr\}\ .
	\end{equation}
	
	\subsection{Dimensional reduction}\label{subsec:action_dimred}
	As an alternative to the construction via the $\mathcal{N}=\,1$ superfield formalism, an on-shell action of $\mathcal{N}=\,4$ $D=4$ SYM can be obtained by dimensional reduction from $\mathcal{N}=\,1$ $D=10$ SYM \cite{BSS}
	\begin{equation}
		S^{(10)}\=\sfrac{1}{g^2}\int \drm^{10}x\Bigl\{-\sfrac{1}{4}F^{\Sigma\Theta}F_{\Sigma\Theta}-\sfrac{\im}{2}\blambda\ \Gamma^{\Sigma}\ \Diff_\Sigma\ \lambda\Bigr\}\ ,
	\end{equation}
	where capital greek indices label the ten-dimensional representation of the Lorentz group and $\Gamma^{\Sigma}$ are the ten-dimensional gamma matrices.
	We reduce the gauge field as
	\begin{equation}
		A_\Sigma\=(A_\mu,\ \varphi_i)\ ,
	\end{equation}
	which leads to the reduction of the Yang--Mills term
	\begin{equation}
		-\sfrac14 F^{\Sigma\Theta}F_{\Sigma\Theta} \quad \longrightarrow \quad -\sfrac14 F_{\mu\nu}F^{\mu\nu}-\sfrac12\Diff_\mu\varphi_i\Diff^\mu\varphi_i-\sfrac{1}{4}(\varphi_i\t\varphi_j)^2\ .
	\end{equation}
	For the Dirac term, we write the gamma matrices as
	\begin{equation}\label{eq:Gamma_rep}
		\Gamma^{\mu}\= \unity_8 \otimes \gamma^\mu \ ,\qquad \Gamma^{AB}\= \begin{pmatrix} 0 & \rho^{AB}\\ \rho_{AB} & 0
		\end{pmatrix} \otimes \im\gamma_5\ ,\qquad A,B\=1,2,3,4\ ,
	\end{equation}
	with antisymmetric $4\t 4$-matrices
	\begin{equation}
		(\rho^{AB})_{CD}\=\delta_{AC}\delta_{BD}-\delta_{AD}\delta_{BC}\ ,\qquad (\rho_{AB})_{CD}\=\sfrac{1}{2}\epsilon_{ABFG}(\rho^{FG})_{CD}\=\epsilon_{ABCD}\ .
	\end{equation}
	It is convenient to define antisymmetric
	\begin{equation}\label{eq:varphi_rep}
		\varphi_{I4}\=\sfrac12(\varphi_I+\im\varphi_{I+3})\ ,\qquad \varphi^{AB}\=\sfrac12 \epsilon^{ABCD}\varphi_{CD}\=(\varphi_{AB})^*\ .
	\end{equation}
	In the literature one often finds a factor $\sfrac{1}{\sqrt{2}}$ in front of the first equation in \eqref{eq:varphi_rep}. In order to match the coefficients from our previous analysis, we prefer to chose a normalization $\sfrac12$ instead. Note that this explicitly determines the Clebsch-Gordon coefficients $t\indices{^i_{AB}}$ from \eqref{eq:Clebsch-Gordon} to be
	\begin{equation}\label{eq:CG_coeff}
		\begin{aligned}
			&(t^I)_{J4}\=\sfrac12 \delta_{IJ}\=(t^I)^{J4}\ ,&&(t^{I{+}3})_{J4}\=\sfrac{\im}{2} \delta_{IJ}\=-(t^{I{+}3})^{J4}\ ,\\
			&(t^I)_{JK}\=\sfrac12 \epsilon_{IJK}\=(t^I)^{JK}\ ,&&(t^{I{+}3})_{JK}\=-\sfrac{\im}{2} \epsilon_{IJK}\=-(t^{I{+}3})^{JK}\ .
		\end{aligned}
	\end{equation}
	For a matching of the bosonic and fermionic degrees of freedom, the spinor $\lambda$ has to be a Majorana-Weyl-spinor, which can be realized in the structure
	\begin{equation}\label{eq:spinor_red}
		\lambda\=(\chP^+\chi^1, \ldots,\chP^+\chi^4,\chP^-\tilde{\chi}_1,\ldots, \chP^-\tilde{\chi}_4)^{\mathrm{T}}\ ,\quad \text{with}\quad \tilde{\chi}_A\=\mathrm{C}\bar{\chi}^{A\;\mathrm{T}}\ ,
	\end{equation}
	where $\mathrm{C}$ is the charge conjugation operator in four dimensions. We find that the Dirac term becomes
	\begin{equation}
		-\sfrac{\im}{2}\blambda\ \Gamma^{\Sigma}\ \Diff_\Sigma\ \lambda\quad \longrightarrow\quad -\sfrac{\im}{2}\bpsi_A\ \slashed{\scrD}\indices{^A_B}\ \psi^B\ ,
	\end{equation}
	with the shorthand \eqref{eq:slscrD}, the Majorana-spinors \eqref{eq:Majorana_red} and
	\begin{equation}
		\Phi\indices{^A_B}\=(c^i)\indices{^A_B}\varphi_i\=\bigl[(\rho^{CD})_{AB}\chP^+-(\rho_{CD})_{AB}\chP^-\bigr]\varphi_{CD}\=2\bigl[t\indices{^i_{AB}}\chP^+-t\indices{^{iAB}}\chP^-\bigr]\varphi_i\ .
	\end{equation}
	It is easy to verify that the coefficients $(c^i)\indices{^A_B}$ defined this way are equivalent to those found in the superfield formalism \eqref{eq:cs}.
	
	\section{Coupling flow operator}\label{sec:roperator}
	In this section we construct the coupling flow operator \eqref{eq:def_cflow} first via the canonical construction \ref{subsec:cconstr_rop} and then via dimensional reduction \ref{subsec:cflow_dimred}. Reconciling the two approaches, we propose a unified framework for the operator in \ref{subsec:geometry}.
	
	\subsection{From the canonical construction}\label{subsec:cconstr_rop}
	The basic procedure of the following construction is exactly the same as in the $\mathcal{N}=\,1$ case (see e.g.~\cite{LR2}). We merely have more fields to take into account.
	In order to fix the redundant degrees of freedom of the gauge theory, following the Faddeev-Popov procedure we add a gauge fixing term $S_{\mathrm{gf}}$ with gauge fixing function $\G(\tscrA)$ and ghost fields $\tbC,\ \tC$ to the full action $S_{\textsc{SUSY}}$. We use the on-shell invariant action \eqref{eq:Sinv_6real_on}, so that
	\begin{equation}
		\begin{aligned}
			&S_{\textsc{SUSY}}[\tA, \tvarphi, \tpsi, \tbpsi, \tC, \tbC]\=S_{\mathrm{inv}}[\tA, \tvarphi, \tpsi, \tbpsi]\ +\ S_{\mathrm{gf}}[\tA, \tvarphi, \tC, \tbC]\ ,\\[6pt]
			&S_{\mathrm{inv}}\=\sfrac{1}{g^2}\intdx\;  \Bigl\{-\sfrac{1}{4}\tF^{\mu\nu}\tF_{\mu\nu}-\sfrac12\tDiff_\mu\tvarphi_i\tDiff^\mu\tvarphi_i-\sfrac{\im}{2}\tbpsi_A\stscrD{\indices{^A_B}}\tpsi^B-\sfrac{1}{4}(\tvarphi_i\t\tvarphi_j)^2\Bigr\}\ ,\\[6pt]
			&S_{\mathrm{gf}}\=\sfrac{1}{g^2}\intdx\;\Bigl\{-\sfrac{1}{2\xi} \G(\tA,\tvarphi)^2+g\,\tbC \sfrac{\partial \G(\tA,\tvarphi)}{\partial \tA_\mu}\trD_\mu \tC+g\,\tbC \sfrac{\partial \G(\tA,\tvarphi)}{\partial \tvarphi_i}\tvarphi_i\t \tC\Bigr\}\ ,\\
		\end{aligned}
	\end{equation}
	where we emphasize that the fields are in the (canonical or geometric) scaling where the coupling only occurs as an overall factor $1/g^2$ (and one factor of $g$ multiplying two ghost terms) by explicitly putting tildes on all scaled quantities. The usual dependence on the coupling is recovered after rescaling all fields with an appropriate power of $g$ (i.e.~$\tA=gA$, $\tvarphi=g\varphi$).
	In this scaling we can write the $g$ derivative of the action as a supervariation up to a Slavnov variation\footnote{strictly, this should be written with the auxiliary fields still present, but since we can integrate them out after the construction, we leave them implicit here.},
	\begin{equation}
		\partial_g S_{\textsc{SUSY}}\=-\sfrac{1}{g^3}\bigr\{\delta_\alpha \Delta_\alpha-\sqrt{g}\,s\,\Delta_{\mathrm{gh}}\bigr\}\ ,
	\end{equation}
	with the superfield component
	\begin{equation}
		\Delta_\alpha\=\sfrac14\intdx\Bigl\{-\sfrac12 \tF_{\mu\nu}\gamma^{\mu\nu}\tlambda-(\tPhi\indices{^4_A})^{\+}\slashed{\tscrD}\indices{^A_B}\tpsi^B+\sfrac12(\tPhi\indices{^4_A})^{\+}\tPhi\indices{^A_B}{\t}\tpsi^B\Bigr\}\ ,
	\end{equation}
	the ghost contribution
	\begin{equation}\label{eq:Delta_gh}
		\Delta_{\mathrm{gh}}\=\intdx\;  \Bigl\{\tbC\ \G(\tA, \tvarphi)\Bigr\}\ ,
	\end{equation}
	the supervariations \eqref{eq:supervariations} and the BRST (or Slavnov) variations
	\begin{equation}\label{eq:slavnov_variations}
		\begin{aligned}
			&s\tA_\mu\=\sqrt{g}\;\trD_\mu \tC\ ,\qquad&&s\tlambda\=\sqrt{g}\;\tlambda\times \tC \ ,\qquad&&s\tblambda\=\sqrt{g} \;\tblambda\times \tC \ ,\qquad&\\[4pt]
			&s\tD\=\sqrt{g}\;\tD\times \tC\ ,&&s\tC\=-\sfrac{\sqrt{g}}{2} \;\tC\times \tC\ ,&&s\tbC\=\sfrac{1}{\sqrt{g}}\sfrac{1}{\xi}\;\G(\tA,\tvarphi)\ ,\\
			&s\tvarphi_i\=\sqrt{g}\;\tvarphi_i\t \tC\ ,&&s\tpsi_I\=\sqrt{g} \;\tpsi_I\times \tC \ ,&&s\tFaux\=\sqrt{g} \;\tFaux_I\times \tC\ .&
		\end{aligned}
	\end{equation}
	An intermediate scaled coupling flow operator is given by \cite{L1,DL1}
	\begin{equation} \label{eq:gaugeR}
		\R[\tscrA] \= -\im\,\bcontraction{}{\Delta}{_\alpha[\tscrA]\ }{\delta} \Delta_\alpha[\tscrA]\ \delta_\alpha 
		+\sfrac{\im}{\sqrt{g}}\,\bcontraction{}{\Delta}{_{\textrm{gh}}[\tscrA]\ }{s} \Delta_{\textrm{gh}}[\tscrA]\ s
		-\sfrac{1}{\sqrt{g}}\,\bcontraction{}{\Delta}{_\alpha[\tscrA]\ \bigl(}{\delta}  \Delta_\alpha[\tscrA]\ \bigl(\delta_\alpha 
		\bcontraction{}{\Delta}{_{\textrm{gh}}[\tscrA]\bigr)\ }{s} \Delta_{\textrm{gh}}[\tscrA]\bigr)\ s\ ,
	\end{equation}
	where we introduced the shorthand $\tscrA=(\A_\mu,\ \tvarphi_i)$ and the contractions indicate either gaugino or ghost propagators.\footnote{Note that the coupling flow acts on observables $X[\tscrA]$ and $\delta_\alpha \tscrA_\Sigma$ contains a gaugino field, whereas $s\;\tscrA_\Sigma$ contains a ghost field.} The calculation can be found in Appendix \ref{app:cconstr}. After rescaling the fields $\tscrA=g\scrA$ according to the scheme developed in \cite{LR1}, the final expression for the coupling flow operator is
	\begin{equation}\label{eq:cf_sf}
		\stackrel{\leftarrow}{R_g}[\scrA]\=\sfrac18\stackrel{\longleftarrow}{\sfrac{\delta}{\delta\scrA_\Gamma}}P\indices{_\Gamma^\Sigma}\ \tr\bigl\{(\scrC_{\Sigma})\indices{^4_A}S\indices{^A_B}\slashed{\scrA}{\indices{^B_C}}\times \slashed{\scrA}{\indices{^{*C}_4}}\bigr\}
		+ \stackrel{\longleftarrow}{\sfrac{\delta}{\delta\scrA_\Gamma}}\Pi\indices{_\Gamma^\Sigma}\scrA_\Sigma G\sfrac{\partial\G(\scrA)}{\partial A_{\nu}}\AL_\nu\ ,
	\end{equation}
	acting to the left to comply with the implicit color index and position argument structure (adopting the notation in section 4 of \cite{ALMNPP}). Note that this has the exact same structure as the result for $\mathcal{N}=\,1$ $D=4$ SYM \cite{LR2}, only with the additional R-symmetry indices. We now give a detailed account of the various quantities involved. First of all, an object that appears very frequently is
	\begin{equation}\label{eq:def_scrC}
		(\scrC_\Sigma)\indices{^A_B}\=\left\{\begin{array}{ll}
			\delta\indices{^A_B}\gamma_\mu & \text{for}\quad \Sigma\=\mu\=0,1,2,3\\
			(c^i)\indices{^A_B} & \text{for}\quad \Sigma\=3+i\=4,5,...,9
		\end{array}\right.\ .
	\end{equation}
	With the shorthands
	\begin{equation}\label{eq:def_gaugino_prop}
		\scrD_\Gamma\=(\Diff_\mu\;,\ g\varphi_i\t\ )\ ,\qquad \slashed{\scrD}{\indices{^A_B}}\=\scrD^\Sigma(\scrC_\Sigma)\indices{^A_B}\=\slashed{\Diff}\delta{\indices{^A_B}}+g\Phi{\indices{^A_B}}\t\ \ ,
	\end{equation}
	we can compactly express the gaugino and ghost propagators $S\indices{^A_B}$, $G$, defined by
	\begin{equation}
		\bcontraction{}{\psi}{^A(x)}{\bpsi}\psi^A(x)\bpsi_B(y)\=-S\indices{^A_B}(x,y;\scrA)\ ,\qquad 
		\sscrD{\indices{^A_C}}\;S\indices{^C_B}(x,y;\scrA)\=\delta\indices{^A_B}\delta(x-y)\ ,
	\end{equation}
	and
	\begin{equation}
		\im\bcontraction{}{C}{(x)}{\bC}C(x)\bC(y)\=G(x,y;\scrA)\ ,\qquad
		\sfrac{\partial\G(\scrA)}{\partial \scrA_\Gamma}\scrD_\Gamma\;G(x,y;\scrA)\=\delta(x-y)\ ,
	\end{equation}
	respectively. Further we have
	\begin{equation}\label{eq:def_slscrA}
		\slashed{\scrA}{\indices{^A_B}}\=\scrA^\Sigma(\scrC_\Sigma)\indices{^A_B}\=\slA\delta{\indices{^A_B}}+\Phi{\indices{^A_B}}\ ,\qquad \slashed{\scrA}^*{\indices{^A_B}}\ :=\ \slA^*\delta{\indices{^A_B}}+(\Phi{\indices{^A_B}})^{\+}\ ,
	\end{equation}
	where $A^*$ is the conjugate gauge field \eqref{eq:conjugate_fields}. The second term in \eqref{eq:cf_sf} also explicitly contains the longitudinal part of the gauge field $A_\mu^{\mathrm{L}}$ \eqref{eq:split_T_L}. Lastly, we have the natural generalization (c.f.~\cite{LR2}) of the covariant projector
	\begin{equation}\label{eq:cov_proj}
		P\indices{_\Gamma^\Sigma}\=\delta\indices{_\Gamma^\Sigma}-\scrD_\Gamma G \sfrac{\partial \G(\scrA)}{\partial \scrA_\Sigma}\ ,
	\end{equation}
	and its free version
	\begin{equation}
		\Pi\indices{_\Gamma^\Sigma}\=P\indices{_\Gamma^\Sigma}\bigr|_{g=0}\ .
	\end{equation}

	\subsection{From dimensional reduction}\label{subsec:cflow_dimred}
	Since the ten-dimensional theory does not have an off-shell formalism, the operator cannot be constructed canonically in any gauge. However, for the Landau gauge, an expression for $R_g$ was derived and shown to satisfy all necessary conditions in all the critical \cite{BSS} dimensions $D=3,4,6,10$ \cite{ALMNPP}. Hence, in the following we restrict to the Landau gauge hypersurface, so there will be no distinction between $A$ and $A^*$ since $\AL=0$. We start with the formula for the coupling flow operator in $\mathcal{N}=\,1$ $D=10$ SYM:
	\begin{equation}\label{eq:cflow_10}
		R_g[\scrA]\=\sfrac{1}{32}\stackrel{\longleftarrow}{\sfrac{\delta}{\delta\scrA_\Gamma}}{P^{(10)}}\indices{_\Gamma^\Sigma}\ \tr^{(32)}\Bigl\{\Gamma_{\Sigma }\Sten\slashed{\scrA}\times \slashed{\scrA}\Bigl\}\ ,
	\end{equation}
	where the trace is over $32\t 32$ spinor space and
	\begin{equation}
		\slashed{\scrA}\=\Gamma^{\Sigma }\scrA_\Sigma\ .
	\end{equation}
	We now apply the same scheme for dimensional reduction that we have used to reduce the action in Section \ref{subsec:action_dimred}.
	It is easy to establish that ${P^{(10)}}\indices{_\Gamma^\Sigma}\longrightarrow {P^{(4)}}\indices{_\Gamma^\Sigma}$ under dimensional reduction from ten to four dimensions. We can relate $\Sten$ and $\Sfour\equiv S$ by dimensional reduction. When leaving out the superscript indicating the number of dimensions, we always mean the four-dimensional object. In ten dimensions, the $32\t 32$ matrix $\Sten$ is given by the contraction
	\begin{equation}
		\Sten\=-\bcontraction{}{\lambda}{}{\blambda}\lambda\blambda\ .
	\end{equation}
	With the dimensional reduction of the spinor $\lambda$ \eqref{eq:spinor_red} and our definition of the four-dimensional Majorana spinors \eqref{eq:Majorana_red}, we can decompose the $32\t 32$ matrix as
	\begin{equation}\label{eq:Sten_block}
		\Sten \= \left(\begin{array}{c|c}
			(\chP^+S{\indices{^A_B}}\chP^-)_{\alpha\beta} & (\chP^+S{\indices{^{AB}}}\chP^+)_{\alpha\beta} \\
			\hline
			(\chP^-S{\indices{_{AB}}}\chP^-)_{\alpha\beta} & (\chP^-S{\indices{_A^B}}\chP^+)_{\alpha\beta}
		\end{array}\right)\ ,
	\end{equation}
	where each block is a $16\t 16$ matrix with `inner' indices $A,\ B$ and `outer' indices $\alpha,\ \beta$, all ranging from one to four. Due to the chiral projectors, the position of the indices $A,\ B$ matches the R-symmetry transformation properties, i.e.~upper indices transform as a $\mathbf{4}$ and lower indices as a $\bar{\mathbf{4}}$.
	We can interpret $\Sten$ as an $8\t 8$ matrix with $4\t 4$ matrix-valued entries and take a partial trace in the $8\t 8$ matrix space so that we are left with a trace over $4\t 4$ matrices (over the outer indices). 
	Using the representation of Gamma matrices \eqref{eq:Gamma_rep} and bosonic fields \eqref{eq:varphi_rep}, we further write $\slashed{\scrA}$ in the same block notation as
	\begin{equation}
			\slashed{\scrA}\=\Gamma^{\Sigma }\scrA_\Sigma\=\left(\begin{array}{c|c}
				\slA_{\alpha\beta}\delta\indices{^A_B} & (\im\gamma_5)_{\alpha\beta}\varphi^{AB} \\
				\hline
				(\im\gamma_5)_{\alpha\beta}\varphi_{AB} & \slA_{\alpha\beta}\delta\indices{_A^B}
			\end{array}\right)\ .
	\end{equation}
	When multiplying two block matrices, we simply have to contract inner with inner indices and outer with outer indices.
	This leads to
	\begin{equation}
		\slashed{\scrA}\t\slashed{\scrA}\=\left(\begin{array}{c|c}
			(\slA\t\slA)_{\alpha\beta}\delta\indices{^A_B}+(\unity_4)_{\alpha\beta}\varphi^{AC}{\t}\varphi_{CB} & 2(\slA\im\gamma_5)_{\alpha\beta}\t\varphi^{AB} \\
			\hline
			2(\im\gamma_5\slA)_{\alpha\beta}\t\varphi_{AB} & (\slA\t\slA)_{\alpha\beta}\delta\indices{_A^B}+(\unity_4)_{\alpha\beta}\varphi_{AC}{\t}\varphi^{CB}
		\end{array}\right)\ .
	\end{equation}
	In order to perform the partial trace, one multiplies the block matrices in \eqref{eq:cflow_10} and then takes the trace over the blocks. For example, a simple contribution would be
	\begin{equation}
		\begin{aligned}
			&\tr^{\mathrm{(32)}}\ \Bigl\{\Gamma_\mu \ \Sten\ \unity_8\otimes (\slA\t\slA)\Bigr\}
			\\
			&\=\tr^{\mathrm{(32)}}\ \left(\begin{array}{c|c}
				(\gamma_{\mu})_{\alpha\gamma}(\chP^+S\indices{^A_B}\chP^-)_{\gamma\delta}(\slA\t\slA)_{\delta\beta} & (\gamma_{\mu})_{\alpha\gamma}(\chP^+S^{AB}\chP^+)_{\gamma\delta}(\slA\t\slA)_{\delta\beta} \\
				\hline
				(\gamma_{\mu})_{\alpha\gamma}(\chP^-S_{AB}\chP^-)_{\gamma\delta}(\slA\t\slA)_{\delta\beta} & (\gamma_{\mu})_{\alpha\gamma}(\chP^-S\indices{_A^B}\chP^+)_{\gamma\delta}(\slA\t\slA)_{\delta\beta}
			\end{array}\right)\\
			&\=\tr^{\mathrm{(4)}}\bigl\{\gamma_\mu \chP^+S\indices{^A_A}\chP^- \slA\t\slA\bigr\}\ +\ \tr^{\mathrm{(4)}}\bigl\{\gamma_\mu \chP^-S\indices{_A^A}\chP^+ \slA\t\slA\bigr\}
			\=\tr^{\mathrm{(4)}}\bigl\{\gamma_\mu S\indices{^A_A}\slA\t\slA\bigr\}\ ,
		\end{aligned}
	\end{equation}
	where in the last step we used the cyclicity of the trace to commute the chiral projectors. In the last step, since the positions of the indices of the gaugino propagator in the first vs.~second term do not match up, the R-symmetry transformation properties become slightly nontransparent. This is a consequence of the way we defined our Majorana spinors \eqref{eq:Majorana_red}. Step by step, one establishes the relation
	\begin{equation}\label{eq:trace_reduction}
		\tr^{(32)}\Bigl\{\Gamma_{\Sigma }\Sten\slashed{\scrA}\times \slashed{\scrA}\Bigl\}
			\=\tr^{(4)}\Bigl\{({\scrC_{\Sigma}})\indices{^A_B}S{\indices{^B_C}}\slashed{\scrA}{\indices{^C_D}}\times \slashed{\scrA}^*{\indices{^D_A}}\Bigl\}\ ,
	\end{equation}
	with the same definitions \eqref{eq:def_scrC}, \eqref{eq:def_gaugino_prop}, \eqref{eq:def_slscrA} as in the canonical construction.
	Although the R-symmetry indices in \eqref{eq:trace_reduction} cannot be strictly assigned to $\mathbf{4}$'s or $\bar{\mathbf{4}}$'s, by construction through the above dimensional reduction, this object is invariant under R-symmetry transformations for $\Sigma=\mu$ and transforms as a $\mathbf{6}$ of $\SU(4)$ for $\Sigma=3+i$.\footnote{This can also be verified through explicit calculation by splitting the quantities in \eqref{eq:trace_reduction} into their chiral contributions that have fixed transformation properties. Due to the `hidden' chiral projectors, all the contributions that do not transform appropriately are projected out of the trace.} 
	
	In total, we have shown that the coupling flow operator obtained from dimensional reduction on the Landau gauge hypersurface is
	\begin{equation}\label{eq:cf_dr}
		\stackrel{\leftarrow}{R_g}[\scrA]\= \sfrac{1}{32}\stackrel{\longleftarrow}{\sfrac{\delta}{\delta\scrA_\Gamma}}{P}\indices{_\Gamma^\Sigma}\ \tr\Bigl\{(\scrC_{\Sigma})\indices{^A_B}S\indices{^B_C}\slashed{\scrA}{\indices{^C_D}}\times \slashed{\scrA}^*{\indices{^{D}_A}}\Bigl\}\ .
	\end{equation}
	\subsection{Unified R-symmetric framework}\label{subsec:geometry}
	The goal of this subsection is to reconcile the two results \eqref{eq:cf_sf} and \eqref{eq:cf_dr}, since they are clearly not identical. For simplicity, we for now restrict to the Landau gauge, for which the two results read
	\begin{align}
			&\stackrel{\leftarrow}{R_g}[\scrA]\= \sfrac{1}{8}\stackrel{\longleftarrow}{\sfrac{\delta}{\delta\scrA_\Gamma}}{P}\indices{_\Gamma^\Sigma}\ \tr\Bigl\{(\scrC_{\Sigma})\indices{^4_B}S\indices{^B_C}\slashed{\scrA}{\indices{^C_D}}\times \slashed{\scrA}^*{\indices{^{D}_4}}\Bigl\}&&\text{from canonical construction}\ ,\label{eq:geometric_case1}\\
			&\stackrel{\leftarrow}{R_g}[\scrA]\= \sfrac{1}{32}\stackrel{\longleftarrow}{\sfrac{\delta}{\delta\scrA_\Gamma}}{P}\indices{_\Gamma^\Sigma}\ \tr\Bigl\{(\scrC_{\Sigma})\indices{^A_B}S\indices{^B_C}\slashed{\scrA}{\indices{^C_D}}\times \slashed{\scrA}^*{\indices{^{D}_A}}\Bigl\}&&\text{from dimensional reduction}\ .\label{eq:geometric_case2}
	\end{align}
	We begin this discussion with two universal observations. Firstly, the general definition of the coupling flow operator $R_g[\phi]$ \eqref{eq:def_cflow} shows that it maps real observables to other real observables.\footnote{That is, modulo imaginary terms that vanish when taking the expectation value. For simplicity, we ignore the possibility of such extra imaginary terms.} Hence, we require the kernel $K$ of the coupling flow operator
	\begin{equation}
		R_g[\phi] \= \int\!\diff x\ K[\phi;\;x]\,\frac{\delta}{\delta\phi(x)}
	\end{equation}
	to be real. Note that for our cases and notation above, the kernel carries a ten-dimensional index $\Gamma$, so that
	\begin{equation}
		\stackrel{\leftarrow}{R_g}[\scrA]\=\stackrel{\longleftarrow}{\sfrac{\delta}{\delta\scrA_\Gamma}}K_\Gamma\ ,
	\end{equation}
	with implicit integration. Secondly, we recall the three conditions \eqref{eq:cf_cond12}, \eqref{eq:cf_cond3} for a coupling flow operator in a general gauge theory
	\begin{equation}\label{eq:cf_cond123}
		(\partial_g+R_g[\phi])S^{\mathrm{b}}_g[\phi]\=0\ , \qquad(\partial_g+R_g[\phi])S^{\mathrm{f}}_g[\phi]\=\int\!\diff x\ \frac{\delta K[\phi;\;x]}{\delta \phi(x)}\ ,\qquad(\partial_g+R_g[\phi])\G(\phi)\=0\ .
	\end{equation}
	It is easy to see that given two coupling flow operators $R_g^{(1)}$ and $R_g^{(2)}$, the linear combination
	\begin{equation}
		R'_g\ :=\ pR_g^{(1)} + qR_g^{(2)}\quad \text{with}\quad p,q\ \in\ \mathbb{R}\quad \text{and}\quad p+q\=1
	\end{equation}
	is again a coupling flow operator (i.e.~satisfies \eqref{eq:cf_cond123}). Thus, the coupling flow operator obeys a principle of superposition with real coefficients that add up to one.
	
	The next ingredient is the fact that in the canonical construction we had to single out one of the supersymmetries to obtain a working superfield formalism. In this work, we chose the `fourth' one, resulting in the index 4 at the beginning and the end of the trace in \eqref{eq:geometric_case1}. This choice is of course arbitrary. We could have equivalently chosen any of the three other supersymmetries. This, together with the principle of superposition allows us (in the Landau gauge) to build a more general coupling flow operator by inserting a diagonal matrix with trace four (or as we prefer, the identity plus a traceless matrix $\mathrm{L}$) in R-space in the trace in the kernel. In anticipation of the rest of this discussion, we write down the general ansatz
	\begin{equation}
		\stackrel{\leftarrow}{R_g}[\scrA]\= \sfrac{1}{32}\stackrel{\longleftarrow}{\sfrac{\delta}{\delta\scrA_\Gamma}}{P}\indices{_\Gamma^\Sigma}\ \tr\Bigl\{(\scrC_{\Sigma})\indices{^A_B}S\indices{^B_C}\slashed{\scrA}{\indices{^C_D}}\times \slashed{\scrA}^*{\indices{^{D}_E}}(\delta\indices{^E_A}+\mathrm{L}\indices{^E_A})\Bigl\}\ .\label{eq:geometric_case_general}
	\end{equation}
	The two cases \eqref{eq:geometric_case1}, \eqref{eq:geometric_case2} correspond to
	\begin{equation}
		\mathrm{L}\=\operatorname{diag}(-1,-1,-1,+3)\quad\text{and}\quad \mathrm{L}\=0\ ,
	\end{equation}
	respectively. From the discussion so far we know that we can reach any element of
	\begin{equation}\label{eq:L_starting_point}
		\Bigl\{\mathrm{L}\=\operatorname{diag}(q_1,q_2,q_3,q_4)\quad\text{with}\quad \sum_i q_i\=0\ \Bigr\}\ =:\ \mathfrak{h}\ ,
	\end{equation}
	the Cartan subalgebra of the Lie algebra $\mathfrak{su}(4)$.
	By applying an $\SU(4)$ R transformation on the various factors in \eqref{eq:geometric_case_general}, we can deduce how $\mathrm{L}$ effectively transforms under R transformations. One finds that the underlying structure of the chiral projectors requires that just like the Majorana spinors \eqref{eq:Majorana_red}, $\mathrm{L}$ splits up into two chiral parts
	\begin{equation}
		\mathrm{L}\=L\ \chP^-+L^*\ \chP^+
	\end{equation}
	that are complex conjugate to each other. The matrix $L$ transforms in the adjoint $\mathbf{15}$ of $\SU(4)$
	\begin{equation}\label{eq:L_transformations}
		L\quad\longrightarrow\quad ULU^{\+}\ ,\quad \with U\ \in\ \SU(4)\ .
	\end{equation}
	We notice that \eqref{eq:L_transformations} preserves zero trace and hermiticity $L^{\+}=L$ so that the group action of $\SU(4)$ on $\mathfrak{h}$ \eqref{eq:L_starting_point} generates the entire Lie algebra $\mathfrak{su}(4)$. Any $L\ \in\ \mathfrak{su}(4)$ can be unitarily diagonalized such that we can classify it in terms of its four eigenvalues $q_i$. Invariants under the adjoint action are $\tr\ L^m$ for any integer $m\geq 1$, but with only the first four
	\begin{equation}
		\tr\ L\=\sum q_i\=0\ ,\quad \tr\ L^2\=\sum q_i^2\ ,\quad \tr\ L^3\=\sum q_i^3\ ,\quad \tr\ L^4\=\sum q_i^4
	\end{equation}
	functionally independent. We can equivalently characterize a generic orbit by the eigenvalues $(q_1,\ q_2,\ q_3,\ q_4)$ with $\sum\ q_i=0$ or by $(\tr\ L^2,\  \tr\ L^3,\ \tr\ L^4)$. This gives us three real parameters matching the dimension of the Cartan subalgebra. For a generic $L\ \in\ \mathfrak{su}(4)$ with all eigenvalues $q_i$ distinct, the stabilizer of the adjoint action is the maximal torus $\mathrm{S}(\mathrm{U}(1)^4)\ \cong\ \mathrm{U}(1)^3$ and its orbit under the action is the 12-dimensional flag manifold
	\begin{equation}\label{eq:coset}
		\faktor{\SU(4)}{\mathrm{U}(1)^3}\ .
	\end{equation}
	For singular $L$, i.e.~some eigenvalues $q_i$ coinciding, the stabilizer is larger and the orbit smaller. Table \ref{tab:stab} summarizes all cases. 
	\begin{table}[H]
		\centering
		\setlength\tabcolsep{0pt}
		\caption{Stabilizer subgroups $X$ of $\SU(4)$ acting on $L\ \in\ \mathfrak{su}(4)$, depending on the degeneracy of the eigenvalues $q_i$. The last column indicates the number of free parameters for the coupling flow operator. It is computed by adding the number of degrees of freedom (dofs) in the choice of the $q_i$'s to the dimension of the orbit $\SU(4)/X$.}
		\label{tab:stab}
		\begin{tabular*}{\textwidth}{@{\extracolsep{\fill}} ccccc }
			\toprule
			\multicolumn{1}{c}{Degeneracy} & \multicolumn{1}{c}{dofs} & \multicolumn{1}{c}{Stabilizer $X$} & \multicolumn{1}{c}{$\operatorname{dim}(X)$} & \multicolumn{1}{c}{$\#$ free param.}\\
			\midrule
			all $q_i$ distinct & 3 & $\mathrm{S}(\operatorname{U}(1)^4)$ & 3 & 15 \\
			two $q_i$ equal & 2 & $\mathrm{S}(\operatorname{U}(2)\t\operatorname{U}(1)^2)$ & 5 & 12 \\
			two equal pairs & 1 & $\mathrm{S}(\operatorname{U}(2)\t\operatorname{U}(2))$ & 7 & 9 \\
			three $q_i$ equal & 1 & $\mathrm{S}(\operatorname{U}(3)\t\operatorname{U}(1))$ & 9 & 7 \\
			all $q_i\ =\ 0$ & 0 & $\SU(4)$ & 15 & 0 \\
			\bottomrule
		\end{tabular*}
	\end{table}
	The fully degenerate case corresponds to $L=0$ \eqref{eq:geometric_case2}. It is a fixed point under all $\SU(4)$ transformations. The canonical construction on the other hand led to a configuration where three of the $q_i$'s were equal, so that only an $\mathrm{S}(\operatorname{U}(3)\t\operatorname{U}(1))$ subgroup leaves the configuration invariant. This suggests that for $L$ with three degenerate $q_i$'s, the points in its orbit under the adjoint action of $\SU(4)$ are those that originate from an off-shell formalism and hence allow for an arbitrary choice of the gauge fixing function. For these cases, when choosing gauges other than the Landau gauge, the second term in \eqref{eq:cf_sf} has to be added to the general formula \eqref{eq:geometric_case_general}.
	
	In Appendix \ref{app:conditions}, we directly check that \eqref{eq:geometric_case_general} indeed satisfies the infinitesimal conditions of a coupling flow operator.
	
	\section{Nicolai maps}\label{sec:nmaps}
	In this section we use the notation of \cite{LR2} to write down Nicolai maps to second order. We briefly recall the most important aspects of this compact notation. Derivatives of the scalar propagator $C$ are simply written as indices, so that e.g.~$\partial_\mu\partial_\nu C \equiv C_{\mu\nu}$. Further, all objects are implicitly matrices in color space with the exception that the last quantity in each term is a vector in color space. Moreover, implicit integration kernels are convoluted with insertions of bosonic fields $A$ or $\varphi$.
	
	A Nicolai map for $\mathcal{N}=\,4$ SYM to $\mathcal{O}(g^2)$ in the Landau gauge was already found in \cite{NP} by directly reducing the result from $\mathcal{N}=\,1$ $D=10$ 
	\begin{equation}
		T_g \scrA_\Sigma\=\scrA_\Sigma-gC^\Theta \scrA_\Sigma \scrA_\Theta+\sfrac{3}{2}g^2 C^\Theta \scrA^\Gamma C_{[\Sigma}\scrA_\Theta \scrA_{\Gamma]}+\mathcal{O}(g^3)\ ,
	\end{equation}
	down to four dimensions with the simple prescription $\scrA\=(A_\mu,\ \varphi_i)$ and $\partial_{3+i} \equiv 0$, which leads to
	\begin{equation}\label{eq:NM_dr1}
		\begin{aligned}
			T_gA_\mu\=A_\mu-gC^\rho A_\mu A_\rho +\sfrac{3}{2}g^2 C^\rho A^\lambda C_{[\mu}A_\rho A_{\lambda]}+g^2C^\rho \varphi_i C_{[\mu} A_{\rho]}\varphi_i +\mathcal{O}(g^3)\ ,
		\end{aligned}
	\end{equation}
	\begin{equation}\label{eq:NM_dr2}
		\begin{aligned}
			T_g\varphi_i\=\varphi_i-gC^\rho \varphi_i A_\rho +g^2 C^{[\rho} A^{\lambda]} C_{\lambda}\varphi_i A_{\rho}+\sfrac{1}{2}g^2C^\rho \varphi_j C_\rho \varphi_j\varphi_i+\mathcal{O}(g^3)\ .
		\end{aligned}
	\end{equation}
	In this section, we want to explicitly show the ambiguity of the $\mathcal{N}=\,4$ map by computing (Appendix \ref{app:formulae}) and testing (Appendix \ref{app:tests}) four different maps to second order in the Landau gauge corresponding to the points
	\begin{equation}
		L\=\operatorname{diag}(+3,-1,-1,-1)\ ,\ ...\ ,\ \operatorname{diag}(-1,-1,-1,+3)\ ,
	\end{equation}
	in $\mathfrak{su}(4)$ and denote the respective coupling flow operators as
	\begin{equation}
		\stackrel{\leftarrow}{R_g}{^{(A)}}[\scrA]\= \sfrac{1}{8}\stackrel{\longleftarrow}{\sfrac{\delta}{\delta\scrA_\Gamma}}{P}\indices{_\Gamma^\Sigma}\ \tr\Bigl\{(\scrC_{\Sigma})\indices{^A_B}S\indices{^B_C}\slashed{\scrA}{\indices{^C_D}}\times \slashed{\scrA}^*{\indices{^{D}_A}}\Bigl\}\ .
	\end{equation}
	with no sum over $A=1,2,3,4$. In general \cite{LR1}, the Nicolai map can be obtained from the perturbative expansion
	\begin{equation}\label{eq:series_R}
		R_g[\scrA] \= \sum_{k=1}^\infty g^{k-1} \rR_k[\scrA] \= \rR_1[\scrA] + g\,\rR_2[\scrA] + g^2 \rR_3[\scrA] + \ldots\ ,
	\end{equation}
	via
	\begin{equation}
		\begin{aligned}
			T_g\scrA &\= \scrA \ -\ g\,\rR_1 \scrA \ -\ \sfrac12g^2\bigl(\rR_2-\rR_1^2\bigr)\scrA\ +\ {\cal O}(g^3)\ .
		\end{aligned}
	\end{equation}
	When putting all the contributions together, we find the four distinct Nicolai maps
	\begin{equation}\label{eq:NM_dr1_4}
		\begin{aligned}
			T_g^{(4)} A_\mu \=&A_\mu -gC^\rho A_\mu A_\rho +\sfrac{3}{2}g^2 C^\rho A^\lambda C_{[\mu}A_\rho A_{\lambda]} + g^2C^\rho \varphi_i C_{[\mu}  A_{\rho]}\varphi_i\\
			&\textcolor{blue}{-\sfrac{1}{2}g^2\Pi\indices{_\mu^\nu}\epsilon_{\nu\lambda\rho\sigma}\sum_{J=1}^3[C^\lambda\varphi_J C^\rho \varphi_{J+3}A^\sigma-C^\lambda\varphi_{J+3} C^\rho \varphi_{J}A^\sigma +C^\lambda A^\rho C^\sigma\varphi_{J+3}\varphi_J]}+\mathcal{O}(g^3)\ ,
		\end{aligned}
	\end{equation}
	\begin{equation}\label{eq:NM_dr1_K}
		\begin{aligned}
			T_g^{(K)} A_\mu \=&A_\mu -gC^\rho A_\mu A_\rho +\sfrac{3}{2}g^2 C^\rho A^\lambda C_{[\mu}A_\rho A_{\lambda]} + g^2C^\rho \varphi_i C_{[\mu}  A_{\rho]}\varphi_i\\
			&\textcolor{blue}{+\sfrac{1}{2}g^2\Pi\indices{_\mu^\nu}\epsilon_{\nu\lambda\rho\sigma}\sum_{J=1}^3(-)^{\delta_{KJ}}[C^\lambda\varphi_J C^\rho \varphi_{J+3}A^\sigma-C^\lambda\varphi_{J+3} C^\rho \varphi_{J}A^\sigma +C^\lambda A^\rho C^\sigma\varphi_{J+3}\varphi_J]}+\mathcal{O}(g^3)\ ,
		\end{aligned}
	\end{equation}
	and
	\begin{equation}\label{eq:NM_dr2_4}
		\begin{aligned}
			T_g^{(4)} \varphi_I\=&\varphi_I-gC^\rho \varphi_I A_\rho+g^2 C^{[\rho}A^{\lambda]}C_\lambda \varphi_IA_\rho + \sfrac{1}{2}g^2C^\rho \varphi_j C_\rho  \varphi_j\varphi_I\\
			&\textcolor{blue}{-\sfrac14 g^2 \epsilon_{\mu\nu\rho\lambda}[C^\mu \varphi_{I+3} C^\nu A^\rho A^\lambda + 2C^\mu A^\nu C^\rho \varphi_{I+3}A^\lambda]}\\
			&\textcolor{blue}{-\sfrac12 g^2 C^\rho \sum_{J=1}^3[\varphi_{I+3}C_\rho \varphi_{J+3}\varphi_J+ \varphi_{J}C_\rho \varphi_{I+3}\varphi_{J+3}- \varphi_{J+3}C_\rho \varphi_{I+3}\varphi_{J}]}+\mathcal{O}(g^3)\ ,
		\end{aligned}
	\end{equation}
	\begin{equation}\label{eq:NM_dr2_K}
		\begin{aligned}
			T_g^{(K)} \varphi_I\=&\varphi_I-gC^\rho \varphi_I A_\rho+g^2 C^{[\rho}A^{\lambda]}C_\lambda \varphi_IA_\rho + \sfrac{1}{2}g^2C^\rho \varphi_j C_\rho  \varphi_j\varphi_I\\
			&\textcolor{blue}{+\sfrac14 g^2 \epsilon_{\mu\nu\rho\lambda}(-)^{\delta_{IK}}[C^\mu \varphi_{I+3} C^\nu A^\rho A^\lambda + 2C^\mu A^\nu C^\rho \varphi_{I+3}A^\lambda]}\\
			&\textcolor{blue}{-\sfrac12 g^2 C^\rho(-)^{\delta_{IK}} \sum_{J=1}^3[\varphi_{I+3}C_\rho \varphi_{J+3}\varphi_J+ \varphi_{J}C_\rho \varphi_{I+3}\varphi_{J+3}- \varphi_{J+3}C_\rho \varphi_{I+3}\varphi_{J}]}\\
			&\textcolor{blue}{+ g^2 C^\rho[\varphi_{I+3}C_\rho \varphi_{K+3}\varphi_K+ \varphi_{K}C_\rho \varphi_{I+3}\varphi_{K+3}- \varphi_{K+3}C_\rho \varphi_{I+3}\varphi_{K}]}+\mathcal{O}(g^3)\ ,\\
		\end{aligned}
	\end{equation}
	and the blue parts of $T_g^{(4)} \varphi_{I+3}$, $T_g^{(K)} \varphi_{I+3}$ are given by those of \eqref{eq:NM_dr2_4}, \eqref{eq:NM_dr2_K} with $I$ and $I{+}3$ exchanged and an overall minus sign on the r.h.s., while the black terms are obtained by simply replacing $I$ with $I{+}3$. Note that the black parts of all four maps exactly equal each other and the result from dimensional reduction \eqref{eq:NM_dr1}, \eqref{eq:NM_dr2}, whereas the blue parts differ in signs for the four choices $A=1,2,3,4$. By investigating the explicit contributions (Appendix \ref{app:formulae}) to the coupling flow operators $R_g^{(A)}$, it is easy to check that the symmetric superposition
	\begin{equation}
		R_g\ :=\ \sfrac{1}{4}(R_g^{(1)}+R_g^{(2)}+R_g^{(3)}+R_g^{(4)})\ ,
	\end{equation}
	exactly yields the result from dimensional reduction, as expected. Note that while we can superimpose coupling flow operators, this is not the case for the Nicolai map \eqref{eq:closedT}, since it is not linear in $R_g$. In Appendix \ref{app:tests}, we show through explicit computations that all four maps that we have found in this section indeed satisfy the necessary conditions for a Nicolai map to second order.

	\section{Conclusions and outlook}\label{sec:conclusions}
	In this work we have initiated a systematic study of the Nicolai map in ${\cal N}=\,4$ supersymmetric Yang--Mills theory in four dimensions. Our main result is the explicit form of the coupling flow operator in the Landau gauge $$\stackrel{\leftarrow}{R_g}[\scrA]\= \sfrac{1}{32}\stackrel{\longleftarrow}{\sfrac{\delta}{\delta\scrA_\Gamma}}{P}\indices{_\Gamma^\Sigma}\ \tr\Bigl\{(\scrC_{\Sigma})\indices{^A_B}S\indices{^B_C}\slashed{\scrA}{\indices{^C_D}}\times \slashed{\scrA}^*{\indices{^{D}_E}}(\delta\indices{^E_A}+\mathrm{L}\indices{^E_A})\Bigl\}\ ,$$ with $$\mathrm{L}\=L\ \chP^-+L^*\ \chP^+\ ,$$
	and $L$ any element in the Lie algebra $\mathfrak{su}(4)$. Most importantly, it follows that the operator is subject to a 15-dimensional ambiguity. This can be traced back to the theory-intrinsic $\SU(4)$ R-symmetry. However, by construction, correlators do not depend on the choice of $L$. Building up on previous results in ${\cal N}=\,1$ SYM and with the help of very compact notation, the knowledge of $R_g$ allows for an analogous perturbative construction of the Nicolai map of ${\cal N}=\,4$ SYM via the universal formula $T_g=\overrightarrow{\cal P}\exp\bigl\{-\int_0^g\diff h\,R_h[\scrA]\bigr\}$.
	
	These first steps suggest many future analyses. Natural open questions relate to the effect of including a topological term in the theory, a better understanding of general gauges and a graphical representation for the perturbative expansion. As a potential application of this work, one could compute explicit ${\cal N}=\,4$ quantum correlators $\bigl\langle X[\scrA] \bigr\rangle_g=\bigl\langle X[T_g^{-1}\scrA] \bigr\rangle_0$ with the inverse Nicolai map. Critical future investigations concern how exactly the ambiguity in the coupling flow operator translates to the Nicolai map and, more distantly related to that, whether the framework of the Nicolai map might hint at an integrable structure of ${\cal N}=\,4$ SYM.
	
	\noindent{\bf\large Acknowledgments.\ }
	It is a pleasure to thank Olaf Lechtenfeld for many helpful discussions. We thank the referee for their comments on the initial submission, which have helped to further clarify several important aspects of this paper.
	This work is supported by a PhD grant of the German Academic Scholarship Foundation. 

	\appendix
	\newpage

	\section{Details on the $\mathcal{N}=\,1$ superfield formalism}\label{app:superfields}
	In the conventions from \cite{WB}\footnote{up to a global sign to recover a plus sign in the field strength and covariant derivative instead of a minus sign.} 
	and in the Wess-Zumino (WZ) gauge, the vector superfield ($V^\+=V$) takes the form
	\begin{equation}
		\begin{aligned}
			V&\=\thet\sigma^\mu\bthet A_\mu(x) -\im\thet^2\bthet \blambda(x) +\im \bthet^2\thet\lambda(x)-\sfrac{1}{2} \thet^2\bthet^2\D(x)\\
			&\=\thet\sigma^\mu\bthet A_\mu(y\hphantom{^\+}) -\im\thet^2\bthet \blambda(y\hphantom{^\+}) +\im \bthet^2\thet\lambda(y\hphantom{^\+})-\sfrac{1}{2} \thet^2\bthet^2[\D(y\hphantom{^\+})-\im\Diff^\mu A_\mu(y\hphantom{^\+})]\\
			&\=\thet\sigma^\mu\bthet A_\mu(y^\+) -\im\thet^2\bthet \blambda(y^\+) +\im \bthet^2\thet\lambda(y^\+)-\sfrac{1}{2} \thet^2\bthet^2[\D(y^\+)+\im\Diff^\mu A_\mu(y^\+)]\\
		\end{aligned}
	\end{equation}
	where $y=x+\im\thet\sigma\bthet$ and $y^\+=x-\im\thet\sigma\bthet$ parameterize (anti-)chiral superspace. The advantage of the WZ gauge is that
	\begin{equation}
		V^2=-\sfrac{1}{2}\thet^2\bthet^2 A_\mu A^\mu\ ,
	\end{equation}
	whereas all higher powers vanish such that the power series
	\begin{equation}
		\ep^{2V}=1+2V+2 V^2
	\end{equation}
	truncates at the second order. The non-abelian supersymmetric field strength $W_\alpha$ and its conjugate are given by
	\begin{equation}
		\begin{aligned}
			&W_\alpha\=-\sfrac{1}{4}\bDiff\bDiff \ep^{-2V}\Diff_\alpha \ep^{2V}\=+2\im\lambda_\alpha(y\hphantom{^\+})-2\bigl[\delta\indices{_\alpha^\beta}\D(y\hphantom{^\+})-\im\sigma\indices{^{\mu\nu}_\alpha^\beta}F_{\mu\nu}(y\hphantom{^\+})\bigr]\thet_\beta-2\thet^2\slashed{\Diff}_{\alpha\dalpha}\blambda^\dalpha(y\hphantom{^\+})\ ,\\
			&\bW^\dalpha\=-\sfrac{1}{4}\Diff\Diff \ep^{-2V}\bDiff^\dalpha \ep^{2V}\=-2\im\blambda^\dalpha(y^\+)-2\bigl[\delta\indices{^\dalpha_\dbeta}\D(y^\+)+\im\bsigma\indices{^{\mu\nu}^\dalpha_\dbeta}F_{\mu\nu}(y^\+)\bigr]\bthet^\dbeta+2\bthet^2\slashed{\bDiff}^{\dalpha\alpha}\lambda_\alpha(y^\+)\ ,
		\end{aligned}
	\end{equation}
	in chiral superspace, with the superspace covariant derivatives $\Diff_\alpha, \bDiff_\dalpha$. The chiral superfields ($\bDiff_\dalpha \Phi_I=0$, $\Diff_\alpha \Phi^\+_I=0$) have the simple expansions
	\begin{equation}
		\Phi_I\=\phi_I(y)+\sqrt{2}\thet\psi_I(y)+\thet^2F_I(y)\ ,\qquad \Phi^\+_I\=\phi^\+_I(y^\+)+\sqrt{2}\bthet\bpsi_I(y^\+)+\bthet^2F^\+_I(y^\+)\ ,
	\end{equation}
	in chiral superspace
	and the full superspace expansions
	\begin{equation}
		\begin{aligned}
			\Phi_I&\=\phi_I(x)+\im\thet\sigma^\mu\bthet\partial_\mu \phi_I(x)+\sfrac14 \thet^2\bthet^2 \Box \phi_I(x)+\sqrt{2}\thet\psi_I(x)-\sfrac{\im}{\sqrt{2}}\thet^2\partial_\mu\psi_I(x)\sigma^\mu\bthet+\thet^2F_I(x)\ ,\\
			\Phi^\+_I&\=\phi^\+_I(x)-\im\thet\sigma^\mu\bthet\partial_\mu \phi^\+_I(x)+\sfrac14 \thet^2\bthet^2 \Box \phi^\+_I(x)+\sqrt{2}\bthet\bpsi_I(x)+\sfrac{\im}{\sqrt{2}}\bthet^2\thet \sigma^\mu\partial_\mu\bpsi_I(x)+\bthet^2F^\+_I(x)\ .
		\end{aligned}
	\end{equation}
	For the construction of the coupling flow operator we also need the penultimate components of the various contributions to \eqref{eq:Lag}. For the first part, we find
	\begin{equation}\label{eq:W}
		\begin{aligned}
			\sfrac{1}{4}W^\alpha W_\alpha&=-\lambda^2+\bigl[-2\im\D\lambda-2F_{\mu\nu}\lambda \sigma^{\mu\nu}\bigr]\thet+\bigl[-2\im \lambda\sigma^\mu\Diff_\mu\blambda-\sfrac12 F^{\mu\nu}F_{\mu\nu}+\D^2+\sfrac{\im}{4}F^{\mu\nu}F^{\rho\lambda}\epsilon_{\mu\nu\rho\lambda}\bigr]\thet^2\ ,\\
			\sfrac{1}{4}\bW_\dalpha \bW^\dalpha&=-\blambda^2+\bigl[+2\im\D\blambda-2F_{\mu\nu}\blambda \bsigma^{\mu\nu}\bigr]\bthet+\bigl[+2\im \Diff_\mu\lambda\sigma^\mu\blambda-\sfrac12 F^{\mu\nu}F_{\mu\nu}+\D^2-\sfrac{\im}{4}F^{\mu\nu}F^{\rho\lambda}\epsilon_{\mu\nu\rho\lambda}\bigr]\bthet^2\ .\\
		\end{aligned}
	\end{equation}
	Next, we evaluate
	\begin{equation}\label{eq:Phi_sf}
		\epsilon_{IJK}\ \tr\  \Phi_I[\Phi_J,\Phi_K]\=\im \epsilon_{IJK} f^{abc}\Bigl[\phi^a_I\phi^b_J\phi^c_K+3\sqrt{2}\;\thet\ \psi^a_I\phi^b_J\phi^c_K
		+3\thet^2\bigl(F^a_I \phi^b_J \phi^c_K
		-\phi^a_I \psi^b_J \psi^c_K\bigr)\Bigr]\ ,
	\end{equation}
	and the hermitian conjugate analogously. Lastly, we find
	\begin{equation}\label{eq:kinetic_term}
		\begin{aligned}
			\sfrac{1}{\nc}\tr\ \ep^{-2V}\Phi^\+_I \ep^{2V}\Phi_I\=&\Phi^{a\+}_I\Phi^a_I+\sfrac{2}{\nc}\tr\ [T^a,T^b]T^c\ \Phi^{a\+}_I V^b \Phi^c_I+\sfrac{2}{\nc}\tr\ [T^a,T^b][T^c,T^d]\ \Phi^{a\+}_I V^b V^c \Phi^d_I\\
			\=...\ 
			+&\thet^2\bthet\bigl[-\im\sqrt{2}\bsigma^\mu\psi^a_I\Diff_\mu\phi^{a\+}_I+\sqrt{2}F^a_I\bpsi^a_I+2 f^{abc}\phi^{a\+}_I\blambda^b\phi^c_I\bigr]\\
			+&\bthet^2\thet\bigl[-\im\sqrt{2}\sigma^\mu\bpsi^{a}_I\Diff_\mu\phi^a_I+\sqrt{2}F^{a\+}_I\psi^a_I-2 f^{abc}\phi^{a\+}_I\lambda^b\phi^c_I\bigr]\\
			+&\thet^2\bthet^2\bigl[-\Diff_\mu\phi^{a\+}_I\Diff^\mu\phi^a_I+F^{a\+}_IF^a_I+\im\Diff_\mu\bpsi_I^a\bsigma^\mu\psi_I\\
			&\qquad\qquad-f^{abc}\bigl(\im\phi_I^{a\+}\D^b\phi_I^c-\sqrt{2}\phi^{a\+}_I\lambda^b\psi_I^c+\sqrt{2}\bpsi^a_I\blambda^b\phi^c_I\bigr)\bigr]\\
			&\ +\ \textrm{total derivatives}\ ,
		\end{aligned}
	\end{equation}
	where we have left out terms of power 2 or less in $\thet$ and the traces over the $\operatorname{SU}(\nc)$ generators were evaluated with
	\begin{equation}
		[T^a, T^b]\=\im f^{abc}T^c\ , \qquad \tr\ T^aT^b\=\nc\delta^{ab}\ .
	\end{equation}
	From \eqref{eq:Lag} and \eqref{eq:W}, \eqref{eq:Phi_sf}, \eqref{eq:kinetic_term}, 
	we deduce that in Weyl notation the Lagrangian can be written as
	\begin{equation}
		\begin{aligned}
			g^2 \Lag\=&-\sfrac{1}{4}F^a_{\mu\nu}F^{a\mu\nu}-\im\lambda^a\sigma^\mu\Diff_\mu\blambda^a+\sfrac{1}{2}\D^2
			-\sfrac{1}{\sqrt{2}}\epsilon_{IJK}f^{abc}\bigl(F^a_I\phi_J^b\phi_K^c+F^{a\+}_I\phi_J^{b\+}\phi_K^{c\+}\bigr)\\
			&-\Diff_\mu\phi^{a\+}_I\Diff^\mu\phi^a_I-\im\psi_I^a\sigma^\mu\Diff_\mu\bpsi_I+F^{a\+}_IF^a_I
			+\sfrac{1}{\sqrt{2}}\epsilon_{IJK}f^{abc}\bigl(\phi^a_I\psi^b_J\psi^c_K+\phi^{a\+}_I\bpsi^{b}_J\bpsi^{c}_K\bigr)\\
			&-\sqrt{2}f^{abc}\bigl(\psi_I^a\lambda^b\phi^{c\+}_I+\bpsi^a_I\blambda^b\phi^c_I\bigr)-\im f^{abc}\phi_I^{a\+}\D^b\phi_I^c\ ,
		\end{aligned}
	\end{equation}
	up to total derivatives. From the superspace expansions we read off the supersymmetry transformations
	\begin{equation}
		\begin{aligned}
			&\delta \phi_I\=\sqrt{2}\thet\psi_I\ ,\qquad \delta \psi_I \=\im\sqrt{2}\sigma^\mu\bthet\Diff_\mu\phi_I+\sqrt{2}\thet F_I\ ,\qquad \delta F_I\=\im \sqrt{2}\bthet \bsigma^\mu \Diff_\mu \psi_I -2 \phi_I{\times} \blambda\bthet\ ,\\
			&\delta A^\mu\=-\im \blambda \bsigma^\mu \thet+\im \bthet \bsigma^\mu \lambda\ ,\qquad \delta \lambda\=\sigma^{\mu\nu}\thet F_{\mu\nu}+\im\thet\D\ ,\qquad \delta\D\=-\thet\sigma^\mu\Diff_\mu \blambda-\Diff_\mu\lambda\sigma^\mu\bthet\ .
		\end{aligned}
	\end{equation}
	For convenience, we translate the superfield formalism to a four-component Majorana basis using
	\begin{equation}
		\begin{aligned}
			&\lambda^{(\mathrm{M})}\=\begin{pmatrix}
				\lambda_\alpha \\ \blambda^\dalpha
			\end{pmatrix}\ ,\qquad 
			\blambda^{(\mathrm{M})}\=(
			\lambda^\alpha,\  \blambda_\dalpha)\ ,\qquad
			\alpha\=\begin{pmatrix}
				\thet_\alpha \\ \bthet^\dalpha
			\end{pmatrix}\ ,\qquad 
			\balpha\=(
			\thet^\alpha,\  \bthet_\dalpha)\ \\
			&\gamma_\mu\=\begin{pmatrix}
				0 & \sigma_\mu \\ \bsigma_\mu & 0
			\end{pmatrix}\ ,\qquad
			\gamma_5\=\gamma_0\gamma_1\gamma_2\gamma_3\=\begin{pmatrix}
				-i & 0 \\ 0 & i
			\end{pmatrix}\ ,\qquad \mathrm{etc.}\ ,
		\end{aligned}
	\end{equation}
	so that
	\begin{equation}
		\begin{aligned}
			&\blambda^{(\mathrm{M})} \lambda^{(\mathrm{M})} \= \lambda\lambda + \blambda\blambda\ ,\qquad
			\blambda^{(\mathrm{M})} \im\gamma_5 \lambda^{(\mathrm{M})} \= \lambda\lambda - \blambda\blambda\ ,\\
			&\blambda^{(\mathrm{M})} \gamma^\mu \lambda^{(\mathrm{M})} \= \lambda\sigma^\mu\blambda + \blambda\bsigma^\mu\lambda\=2\lambda\sigma^\mu\blambda\ ,\qquad
			\frac{1}{2}\blambda^{(\mathrm{M})} \gamma^{\mu\nu}\alpha\=\lambda \sigma^{\mu\nu}\thet+\blambda\bsigma^{\mu\nu}\bthet\ ,\qquad \text{etc.}\ ,
		\end{aligned}
	\end{equation}
	where the l.h.s.~are in the four-component Majorana basis and the r.h.s.~are in the two-component Weyl basis. Additionally, we need the chiral projectors
	\begin{equation}
		\chP^{\pm}\=\sfrac{1}{2}(1\pm\im\gamma_5)\ ,\qquad \blambda^{(\mathrm{M})}\chP^{+}\lambda^{(\mathrm{M})}\=\lambda\lambda\ ,\quad \blambda^{(\mathrm{M})}\chP^{-}\lambda^{(\mathrm{M})}\=\blambda\blambda\ .
	\end{equation}
	This leads to the Lagrangian in Majorana notation \eqref{eq:Lag_Maj} (leaving the superscript $^{(\mathrm{M})}$ implicit from now on) and to the penultimate component
	\begin{equation}
		\begin{aligned}
			\Dc\=&\balpha\Bigl\{-\D\gamma_5\lambda-\sfrac12 F_{\mu\nu}\gamma^{\mu\nu}\lambda+2\epsilon_{IJK}f^{abc}\bigl[\chP^+\psi^a_I\phi^b_J\phi^c_K+\chP^-\psi^{a}_I\phi^{b\+}_J\phi^{c\+}_K\bigr]+2\im f^{abc}\gamma_5\phi_I^{a\+}\lambda^b\phi^c_I\\
			&+\im \sqrt{2}\bigl[\gamma^\mu\chP^-\psi^{a}_I\Diff_\mu\phi^a_I+\gamma^\mu\chP^+ \psi^a_I\Diff_\mu\phi^{a\+}_I\bigr]-\sqrt{2}\bigl[\chP^+\psi^a_IF^{a\+}_I+\chP^-\psi^a_IF^a_I\bigr]\Bigr\}\ .
		\end{aligned}
	\end{equation}
	For completeness, we note the following hermiticity properties for Majorana spinors $\chi, \xi$:
	\begin{equation}
		\begin{aligned}
			&\bchi\xi\=\bxi \chi\ ,\qquad
			\bchi\gamma^\mu\xi\=-\bxi \gamma^\mu\chi\ ,\qquad
			\bchi\gamma_5\xi\=\bxi \gamma_5\chi\ ,\qquad
			\bchi\gamma^{\mu}\gamma_5\xi\=\bxi\gamma^{\mu} \gamma_5\chi\ ,\\
			&\bchi\gamma^{\mu\nu}\xi\=-\bxi\gamma^{\mu\nu}\chi\ ,\qquad
			\bchi\gamma^{\mu\nu}\gamma_5\xi\=-\bxi\gamma^{\mu\nu} \gamma_5\chi\ ,\qquad	\bar{\chi}\gamma^{\rho\lambda}\gamma_{\mu}\xi\=\bar{\xi}\gamma_\mu \gamma^{\rho\lambda}\chi\ .
		\end{aligned}
	\end{equation}
	
	\noindent\textbf{Consistency checks.} In order to cross-check the expressions above (in the Majorana basis), we performed three consistency checks. Firstly, due to the superfield structure, the penultimate component $\Dc$ has to generate the Lagrangian via its supervariation up to total derivatives
	\begin{equation}
		\sfrac{1}{4}\delta \Dc\bigr|_{\balpha\alpha}\=g^2\Lag\ ,
	\end{equation}
	with $\balpha(\ldots)\alpha\bigr|_{\balpha\alpha}=-\sfrac{1}{2}\tr (\ldots)$. In practice, this requires the Fierz identity for Majorana spinors
	\begin{equation}\label{eq:Fierz_general}
		4\xi\bchi=-(\bchi\xi)+\gamma_\mu(\bchi\gamma^\mu\xi)+\sfrac{1}{2}\gamma_{\mu\nu}(\bchi\gamma^{\mu\nu}\xi)+\gamma_5\gamma_\mu (\bchi\gamma_5\gamma^\mu \xi)+\gamma_5(\bchi\gamma_5\xi)\ .
	\end{equation}
	A second check is making sure that the Lagrangian transforms as a divergence, i.e.~
	\begin{equation}
		\delta \Lag\=\text{divergence}\ .
	\end{equation}
	A third consistency check is verifying the generation of the susy-algebra
	\begin{equation}\label{eq:susy_algebra}
		\{Q_\alpha, \bar{Q}_\beta\}\=2{(\gamma^\mu)}_{\alpha\beta}P_\mu\=-2\im{(\gamma^\mu)}_{\alpha\beta}\partial_\mu\ ,
	\end{equation}
	up to a gauge transformation. With the supercharges $\delta(...)=\balpha_\alpha Q_\alpha(...)$ this can be evaluated by computing the commutator of two supervariations
	\begin{equation}
		[\delta^{(1)},\delta^{(2)}]=[\balpha_{1\alpha} Q_\alpha, \bar{Q}_\beta\alpha_{2\beta}]=\balpha_{1\alpha} \{Q_\alpha, \bar{Q}_\beta\} \alpha_{2\beta}\ ,
	\end{equation}
	acting on each field. One finds that the susy-algebra reads
	\begin{equation}
		\{Q_\alpha, \bar{Q}_\beta\}=-2\im{(\gamma^\mu)}_{\alpha\beta}\partial_\mu-[\omega, \cdot]_{\alpha\beta}+G_{\alpha\beta}(A)\;,
	\end{equation}
	where $\omega=2\im \slashed{A} $, $[\omega, \cdot]^a=f^{abc}\omega^b(\cdot)^c$ and $G_{\alpha\beta}$ is a gauge transformation $A_\mu\rightarrow A_\mu+\partial_\mu \omega$ as required in the WZ gauge.
	
	\noindent\textbf{Stripping-off the susy parameter.} Lastly, we find it convenient to strip-off the susy parameter by setting $\delta \equiv \delta_\alpha\alpha_\alpha$ and $\Dc \equiv \balpha_\alpha\Dc_\alpha$. This yields the supervariations \eqref{eq:supervariations} and the penultimate component \eqref{eq:Dc_Maj} (up to an overall normalization). We note that the fermionic supervariations have gained an extra minus sign, since
	\begin{equation}
		\bchi \delta \lambda\=\bchi_\beta M_{\beta\alpha}\alpha_\alpha\=\bchi_\beta \delta_\alpha \alpha_\alpha \lambda_\beta\=-\bchi_\beta \delta_\alpha \lambda_\beta \alpha_\alpha \quad \Rightarrow \quad \delta_\alpha \lambda_\beta \=-M_{\beta\alpha}\ ,
	\end{equation}
	with some arbitrary spinor $\bchi$. In general, one has to be careful with sign-flips, because fermionic quantities anti-commute with each other.

	\section{Details on the canonical construction}\label{app:cconstr}
	In this appendix we give the detailed calculation for how to get from \eqref{eq:gaugeR} to \eqref{eq:cf_sf}.
	We need
	\begin{equation}
		\begin{aligned}
			\delta_\alpha X[\tscrA]
			&\=-\im\intdx\Bigl(\tbpsi_4\gamma_\mu\sfrac{\delta}{\delta \A_\mu}+\tbpsi_J (c^i)\indices{^J_4}\sfrac{\delta}{\delta\tvarphi_i}\Bigr)_\alpha X[\tscrA]\\
			&\=-\im\intdx\ \Bigl( \tbpsi_A(\hscrC_\Sigma)\indices{^A_4}\sfrac{\delta}{\delta\tscrA_\Sigma}\Bigr)_\alpha X[\tscrA]\ ,
		\end{aligned}
	\end{equation}
	where we introduced the object
	\begin{equation}
		(\hscrC_\Sigma)\indices{^A_4}\=\left\{
		\begin{array}{ll}
			\delta\indices{^A_4}\gamma_\mu & \text{for}\quad \Sigma\=\mu\=0,1,2,3\\
			(c^i)\indices{^A_4} & \text{for}\quad \Sigma\=3+i\=4,5,...,9
		\end{array}\right.\ ,
	\end{equation}
	with matrix-valued entries. It is defined via
	\begin{equation}\label{eq:def_C}
		\delta_\alpha^{(4)}\tscrA_\Sigma\=-\im (\tbpsi_A(\hscrC_\Sigma)\indices{^A_4})_\alpha\ ,
	\end{equation}
	where the $(4)$ indicates that we have singled out one of the four supersymmetries (the `fourth' one).
	Further we have
	\begin{equation}
		sX[\tscrA]\=\sqrt{g}\intdx\ \tscrD_\Gamma \tC\sfrac{\delta}{\delta \tscrA_\Gamma} X[\tscrA]\ ,
	\end{equation}
	as well as the gaugino and ghost propagators, given by
	\begin{equation}
		\bcontraction{}{\tpsi}{^A(x)}{\tbpsi}\tpsi^A(x)\tbpsi_B(y)\=-\tS\indices{^A_B}(x,y;\tscrA)\ ,\qquad 
		\stscrD{\indices{^A_C}}\;\tS\indices{^C_B}(x,y;\tscrA)\=\delta\indices{^A_B}\delta(x-y)\ ,
	\end{equation}
	and
	\begin{equation}
		\im\bcontraction{}{\tC}{(x)}{\tbC}\tC(x)\tbC(y)\=\tG(x,y;\tscrA)\ ,\qquad
		\sfrac{\partial\G(\tscrA)}{\partial \tscrA_\Gamma}\tscrD_\Gamma\;\tG(x,y;\tscrA)\=\delta(x-y)\ ,
	\end{equation}
	respectively.
	The rescaled coupling flow operator then reads
	\begin{equation}
		\tRl[\tscrA]\=\stackrel{\longleftarrow}{\sfrac{\delta}{\delta\tscrA_\Gamma}}\tP\indices{_\Gamma^\Sigma}\tR_\Sigma\ +\ \stackrel{\longleftarrow}{\sfrac{\delta}{\delta\tscrA_\Gamma}}\tscrD_\Gamma \tG\ \G(\tscrA)\ ,
	\end{equation}
	where we introduced the covariant projector
	\begin{equation}
		\tP\indices{_\Gamma^\Sigma}\=\delta\indices{_\Gamma^\Sigma}-\tscrD_\Gamma \tG \sfrac{\partial \G(\tscrA)}{\partial \tscrA_\Sigma}\ ,
	\end{equation}
	and
	\begin{equation}\label{eq:int_R}
		\tR_{\Sigma}\=-\sfrac{1}{4}\tr\Bigl\{\bigl[\sfrac12 \tF_{\mu\nu}\gamma^{\mu\nu}\tS\indices{^4_C}+(\tPhi\indices{^4_A})^{\+}\slashed{\tscrD}{\indices{^A_B}}\tS\indices{^B_C}-\sfrac12(\tPhi\indices{^4_A})^{\+}\tPhi\indices{^A_B}{\t}\tS\indices{^B_C}\bigr](\hscrC_{\Sigma})\indices{^C_4}\Bigl\}\ ,
	\end{equation}
	where the trace is over Majorana spinor space.
	
	The original (unrescaled) coupling flow operator is given by \cite{LR1}
	\begin{equation}
		R_g[\scrA] \= \sfrac1g\,\bigl( \R[\tscrA] -E \bigr)\qquad \textrm{with}\qquad E \= \tscrA_\Gamma\,\sfrac{\delta}{\delta \tscrA_\Gamma}\ .
	\end{equation}
	To isolate the Euler operator $E$, we need the identities
	\begin{align}
		&\gamma^{\rho\lambda}\tF_{\rho\lambda}\=2\slashed{\trD} \sltA +2\partial\cdot \tA-\sltA\times \sltA\ ,\\
		&\slashed{\tDiff}\tS\indices{^4_C}\=\delta\indices{^4_C}-\tPhi\indices{^4_B}\t\tS\indices{^B_C}\ ,
	\end{align}
	which (next to other contributions) generate the $\tA_{\mu}\sfrac{\delta}{\delta \tA_\mu}$ part of $E$.
	Further we use $\stscrD{\indices{^A_B}}\;\tS\indices{^B_C}=\delta\indices{^A_C}$ in the second term of \eqref{eq:int_R} and
	\begin{equation}
		(\tPhi{\indices{^4_A}})^{\+}(\hscrC_{\Sigma})\indices{^A_4}\=\left\{
		\begin{array}{cl}
			0 & \text{for}\quad \Sigma\=\mu\\
			-\unity_4\tvarphi_I-\gamma_5\tvarphi_{I+3} & \text{for}\quad \Sigma\=3+I\\
			+\gamma_5\tvarphi_I-\unity_4\tvarphi_{I+3} & \text{for}\quad \Sigma\=6+I\\
		\end{array}\right.\ .
	\end{equation}
	With $\tr\ \gamma_5=0$, this gives the second part of the Euler operator.
	Straightforward calculations lead to
	\begin{align}
		\tR_{\Sigma}\=\tscrA_{\Sigma}-\sfrac{1}{4}\tr\Bigl\{(\scrC_{\Sigma })\indices{^4_A}\bigl[\sfrac12\tS\indices{^A_4}(2\partial\cdot \tA-\sltA\times \sltA)- \tS\indices{^A_B}\tPhi\indices{^B_4}\t\sltA
		-\sfrac12\tS\indices{^A_B}\tPhi\indices{^B_C}\t(\tPhi\indices{^C_4})^{\+}\bigr]\Bigl\}\ ,
	\end{align}
	where we flipped the order of the quantities in the trace for a more natural implicit color structure. To do so, we have used that $\tR_{\Sigma}$ is real and identities such as
	\begin{equation}
		\begin{aligned}
			&\bpsi\=\psi^{\+}\;\gamma_0\ ,\qquad (\gamma_0)^2\=\unity_4\ ,\qquad (\tS\indices{^A_B})^{\+}\=\gamma_0\;\tS\indices{^B_A}\;\gamma_0\ ,\qquad
			 (\scrC_{\Sigma})\indices{^4_A}\ :=\ \gamma_0\ ((\hscrC_{\Sigma})\indices{^A_4})^{\+}\ \gamma_0\ ,\\& \gamma_\mu^{\+}\=\gamma_0\; \gamma_\mu\; \gamma_0\ ,\qquad \gamma_5^{\+}\=\gamma_0\; \gamma_5 \gamma_0\=-\gamma_5\; \ ,\qquad \gamma_0\; \tPhi\indices{^A_B}\; \gamma_0\=(\tPhi\indices{^B_A})^{\+}\ .
		\end{aligned}
	\end{equation}
	This leads to
	\begin{equation}
		(\scrC_{\Sigma})\indices{^4_A}\=\left\{
		\begin{array}{ll}
			\delta\indices{^4_A}\gamma_\mu & \text{for}\quad \Sigma\=\mu\=0,1,2,3\\
			(c^i)\indices{^4_A} & \text{for}\quad \Sigma\=3+i\=4,5,...,9
		\end{array}\right.\ .
	\end{equation}
	Since $\tR_\Gamma=\tscrA_\Gamma+...$, the Euler operator conveniently cancels and we find (after inserting $\tscrA=g\scrA$) for any linear gauge
	\begin{equation}
		\stackrel{\leftarrow}{R_g}[\scrA]\=-\sfrac14\stackrel{\longleftarrow}{\sfrac{\delta}{\delta\scrA_\Gamma}}P\indices{_\Gamma^\Sigma}\ \tr\Bigl\{(\scrC_{\Sigma })\indices{^4_A}\bigl[\sfrac12 S\indices{^A_4}(\sfrac2g\partial{\cdot} A-\slA\times \slA)- S\indices{^A_B}\Phi\indices{^B_4}\t\slA
		-\sfrac12 S\indices{^A_B}\Phi\indices{^B_C}\t(\Phi\indices{^C_4})^{\+}\bigr]\Bigl\}\ .
	\end{equation}
	Now that the coupling is restored, with
	\begin{equation}
		\sscrD{\indices{^A_B}}\=\slashed{\Diff}\delta{\indices{^A_B}}+g\Phi{\indices{^A_B}}\t\ ,\qquad \text{with}\qquad\Diff_\mu\=\partial_\mu+gA_\mu\t\ \ ,
	\end{equation}
	and $\sscrD{\indices{^A_C}}S\indices{^C_B}\=\delta\indices{^A_B}$, the fermion propagators can be expanded perturbatively:
	\begin{equation}\label{eq:perturb_S}
		S{\indices{^A_B}}\=S_0\delta{\indices{^A_B}}-gS_0\slashed{\scrA}{\indices{^A_C}}S{\indices{^C_B}}\=\sum_{l=0}^{\infty}\bigl(-gS_0\slashed{\scrA}\bigr)^l{\indices{^A_B}}S_0\ ,
	\end{equation}
	with $S_0=\slashed{\partial}^{\;-1}=-\slashed{\partial}C$ and
	\begin{equation}
		\slashed{\scrA}{\indices{^A_B}}\=\slashed{A}\delta\indices{^A_B}+\Phi\indices{^A_B}\ .
	\end{equation}
	In particular $S{\indices{^A_4}}|_{g=0}=0$, so that the coupling flow operator $R_g$ contains no term of order $\sfrac1g$. We use the same procedure to get rid of the $S_0\sfrac2g\partialA|_{g=0}$ contribution as in the $\mathcal{N}=\,1$ case \cite{LR2} with 
	\begin{equation}
		2S_0\partialA\=-2\slAL\=\slA^*-\slA
	\end{equation} 
	to rewrite the first term:
	\begin{equation}
		\begin{aligned}
			&-\sfrac14\stackrel{\longleftarrow}{\sfrac{\delta}{\delta\scrA_\Gamma}}P\indices{_\Gamma^\Sigma}\ \tr\bigl\{(\scrC_{\Sigma })\indices{^4_A}\bigl[\sfrac12 S\indices{^A_4}(\sfrac2g\partial{\cdot} A-\slA\times \slA)\bigr]\bigr\}\\
			\=&-\sfrac18\stackrel{\longleftarrow}{\sfrac{\delta}{\delta\scrA_\Gamma}}P\indices{_\Gamma^\Sigma}\ \tr\Bigl\{(\scrC_{\Sigma })\indices{^4_A}\Bigl[ \sum_{l=0}^{\infty}\bigl(-gS_0\slashed{\scrA}\bigr)^l{\indices{^A_4}}S_0(\sfrac{2}{g}\partial\cdot A-\slA\times \slA)\Bigr]\Bigr\}\\
			\=&-\sfrac18\stackrel{\longleftarrow}{\sfrac{\delta}{\delta\scrA_\Gamma}}P\indices{_\Gamma^\Sigma}\ \tr\Bigl\{(\scrC_{\Sigma })\indices{^4_A}\Bigl[ \sfrac1g\sum_{l=0}^{\infty}\bigl(-gS_0\slashed{\scrA}\bigr)^l{\indices{^A_4}}S_0 (\slA^*-\slA)-\sum_{l=0}^{\infty}\bigl(-gS_0\slashed{\scrA}\bigr)^l_{A4}S_0\slA\times \slA\Bigr]\Bigr\}\\
			\=&-\sfrac18\stackrel{\longleftarrow}{\sfrac{\delta}{\delta\scrA_\Gamma}}P\indices{_\Gamma^\Sigma}\ \tr\Bigl\{(\scrC_{\Sigma })\indices{^4_A}\Bigl[ \sfrac1g\sum_{l=0}^{\infty}\bigl(-gS_0\slashed{\scrA}\bigr)^l{\indices{^A_4}}S_0 (\slA^*-\slA)-\sum_{l=0}^{\infty}\bigl(-gS_0\slashed{\scrA}\bigr)^l{\indices{^A_B}}S_0(\slashed{\scrA}{\indices{^B_4}}-\Phi{\indices{^B_4}})\times \slA\Bigr]\Bigr\}\\
			\=&-\sfrac18\stackrel{\longleftarrow}{\sfrac{\delta}{\delta\scrA_\Gamma}}P\indices{_\Gamma^\Sigma}\ \tr\Bigl\{(\scrC_{\Sigma })\indices{^4_A}\Bigl[ \sfrac1g\sum_{l=0}^{\infty}\bigl(-gS_0\slashed{\scrA}\bigr)^l{\indices{^A_4}}S_0 (\slA^*-\slA)+\sfrac{1}{g}\sum_{l=1}^{\infty}\bigl(-gS_0\slashed{\scrA}\bigr)^l{\indices{^A_4}}\times \slA+S{\indices{^A_B}}\Phi{\indices{^B_4}}\t\slA\Bigr]\Bigr\}\\
			\=&-\sfrac18\stackrel{\longleftarrow}{\sfrac{\delta}{\delta\scrA_\Gamma}}P\indices{_\Gamma^\Sigma}\ \tr\Bigl\{(\scrC_{\Sigma })\indices{^4_A}\Bigl[ \sfrac1g \delta\indices{^A_4} S_0 (\slA^*-\slA)+\sfrac{1}{g}\sum_{l=1}^{\infty}\bigl(-gS_0\slashed{\scrA}\bigr)^l{\indices{^A_4}}\times \slA^*+S{\indices{^A_B}}\Phi{\indices{^B_4}}\t\slA\Bigr]\Bigr\}\\
			\=&-\sfrac18\stackrel{\longleftarrow}{\sfrac{\delta}{\delta\scrA_\Gamma}}P\indices{_\Gamma^\Sigma}\ \tr\Bigl\{(\scrC_{\Sigma })\indices{^4_A}S{\indices{^A_B}}\Bigl[ -\slashed{\scrA}{\indices{^B_4}}\times \slA^*+\Phi{\indices{^B_4}}\t\slA\Bigr]\Bigr\}-\sfrac{1}{g} \stackrel{\longleftarrow}{\sfrac{\delta}{\delta\scrA_\Gamma}}P\indices{_\Gamma^\nu}\AL_\nu\\
			\=&-\sfrac14\stackrel{\longleftarrow}{\sfrac{\delta}{\delta\scrA_\Gamma}}P\indices{_\Gamma^\Sigma}\ \tr\Bigl\{(\scrC_{\Sigma })\indices{^4_A}S{\indices{^A_B}}\Bigl[ -\sfrac12\slashed{\scrA}{\indices{^B_4}}\times \slA^*+\sfrac12\Phi{\indices{^B_4}}\t\slA\Bigr]\Bigr\}+ \stackrel{\longleftarrow}{\sfrac{\delta}{\delta\scrA_\Gamma}}\Pi\indices{_\Gamma^\Sigma}\scrA_\Sigma G\sfrac{\partial\G(\scrA)}{\partial A_{\nu}}\AL_\nu\ ,\\
		\end{aligned}
	\end{equation}
	containing no term of order $1/g$. Putting everything together we find
	\begin{equation}
		\stackrel{\leftarrow}{R_g}[\scrA]\=\sfrac18\stackrel{\longleftarrow}{\sfrac{\delta}{\delta\scrA_\Gamma}}P\indices{_\Gamma^\Sigma}\ \tr\Bigl\{(\scrC_{\Sigma })\indices{^4_A}S{\indices{^A_B}}\bigl[\slashed{\scrA}{\indices{^B_4}}\times \slA^*+\Phi{\indices{^B_4}}\t\slA + \Phi{\indices{^B_C}}\t\dPhi{\indices{^C_4}}\bigr]\Bigl\}+ \stackrel{\longleftarrow}{\sfrac{\delta}{\delta\scrA_\Gamma}}\Pi\indices{_\Gamma^\Sigma}\scrA_\Sigma G\sfrac{\partial\G(\scrA)}{\partial A_{\nu}}\AL_\nu\ ,
	\end{equation}
	which after defining
	\begin{equation}
		\slashed{\scrA}^*{\indices{^A_B}}\=\slA^*\delta{\indices{^A_B}}+(\Phi{\indices{^A_B}})^{\+}\ ,
	\end{equation}
	takes the simple form \eqref{eq:cf_sf}.
	
	\section{Infinitesimal free action condition}\label{app:conditions}
	In this appendix we present a direct proof that the coupling flow operator in the Landau gauge \eqref{eq:geometric_case_general} satisfies the three infinitesimal conditions \eqref{eq:cf_cond12} \eqref{eq:cf_cond3}. The determinant matching condition follows from the other two conditions and the defining relation \eqref{eq:def_nmap}. The gauge condition \eqref{eq:cf_cond3} follows automatically from the form of the covariant projector \eqref{eq:cov_proj}. Thus, we have only left to show the infinitesimal free action condition
	\begin{equation}\label{eq:ifa}
		(\partial_g+R_g)S_g^{\mathrm{b}}[\scrA]\=0\ .
	\end{equation}
	The basic procedure of the proof is equivalent to the one in A.3 of \cite{ALMNPP} for ${\cal N}=\,1$ SYM in $D=3,4,6,10$, but we have to take into account subtleties coming from the additional degrees of freedom in the ${\cal N}=\,4$ case. We can write the bosonic action as
	\begin{equation}
		S_g^{\mathrm{b}}[\scrA]\=\intdx \bigl\{-\sfrac14 \mathcal{F}^{\Sigma\Theta}\mathcal{F}_{\Sigma\Theta}\bigr\}\ ,
	\end{equation}
	with
	\begin{equation}
		\mathcal{F}_{\Sigma\Theta}\=\partial_\Sigma\scrA_\Theta-\partial_\Theta \scrA_\Sigma +g\scrA_\Sigma\t \scrA_\Theta\ ,\qquad \partial_{3+i}\=0\ ,\qquad \scrA_\mu\=A_\mu\ ,\qquad \scrA_{3+i}\=\varphi_i\ .
	\end{equation}
	From these expressions, it is easy to find
	\begin{equation}
		\partial_g S_g^{\mathrm{b}}\=-\sfrac12 \mathcal{F}^{\Sigma\Theta}\scrA_\Sigma{\t} \scrA_\Theta\und \sfrac{\delta S_g^{\mathrm{b}}}{\scrA_\Sigma}\=\scrD_\Theta\mathcal{F}^{\Theta\Sigma}\ ,
	\end{equation}
	with implicit integration. We first show the statement for the particular choice of the coupling flow operator \eqref{eq:geometric_case1} and afterwards generalize the result to the full Lie algebra $\mathfrak{su}(4)$. Concretely, we first prove that
	\begin{equation}\label{eq:ifa_special_case}
		(\partial_g+R_g)S_g^{\mathrm{b}}[\scrA]\=-\sfrac12 \mathcal{F}^{\Sigma\Theta}\scrA_\Sigma{\t} \scrA_\Theta+\sfrac{1}{8}\scrD_\Theta\mathcal{F}^{\Theta\Sigma}\ \tr\bigl\{(\scrC_{\Sigma})\indices{^4_B}S\indices{^B_C}\slashed{\scrA}{\indices{^C_D}}\times \slashed{\scrA}^*{\indices{^{D}_4}}\bigr\}
	\end{equation}
	vanishes. To do so, we use the identities
	\begin{align}
		&\sfrac{1}{4}\tr\bigl\{(\scrC_{\Sigma})\indices{^4_B}(\scrCb_{\Theta})\indices{^B_C}(\scrC_{\Gamma})\indices{^C_D}(\scrCb_{\Psi})\indices{^D_4}\bigr\}\= \eta_{\Sigma\Psi}\eta_{\Theta\Gamma}-\eta_{\Sigma\Gamma}\eta_{\Theta\Psi}+\eta_{\Sigma\Theta}\eta_{\Gamma\Psi}\ ,\label{eq:id_fa_1}\\
		&(\scrC^{\Gamma})\indices{^A_B}\scrD_\Gamma\ S\indices{^B_C}\=\slashed{\scrD}\indices{^A_B} \ S\indices{^B_C}\=\delta\indices{^A_C}\ ,\label{eq:id_fa_2}\\
		&(\scrC_{[\Sigma})\indices{^A_B}(\scrCb_{\Theta]})\indices{^B_C}(\scrC_{\Gamma})\indices{^C_D}\= -2(\scrC_{[\Sigma})\indices{^A_D}\ \eta_{\Theta]\Gamma}+(\scrC_{[\Sigma})\indices{^A_B}(\scrCb_{\Theta})\indices{^B_C}(\scrC_{\Gamma]})\indices{^C_D}\ ,\label{eq:id_fa_3}
	\end{align}
	that are similar to the ones used in \cite{ALMNPP}. Here we have introduced a `conjugate' $\scrCb$ (in the Landau gauge), so that
	\begin{equation}
		\begin{aligned}
			&\scrC_{\mu}\=\unity_4\gamma_\mu\ ,\qquad \scrC_{3+i}\=2[(t^i)^*\chP^+-t^i\chP^-]\ ,&& \slashed{\scrA}{\indices{^A_B}}\=\scrA^\Gamma(\scrC_{\Gamma})\indices{^A_B}\=\slA+\Phi\indices{^A_B}\\
			&\scrCb_{\mu}\=\unity_4\gamma_\mu\ ,\qquad \scrCb_{3+i}\=2[t^i\chP^+-(t^i)^*\chP^-]\ ,&& \slashed{\scrA}^*{\indices{^A_B}}\=\scrA^\Gamma(\scrCb_{\Gamma})\indices{^A_B}\=\slA+(\Phi\indices{^A_B})^{\+}\ ,
		\end{aligned}
	\end{equation}
	with the Clebsch-Gordon coefficients $t^i_{AB}$ as matrices in R-space.
	It should be noted that \eqref{eq:id_fa_1} is only valid up to terms that vanish when contracted with fields in the adjoint representation of the gauge group due to the Jacobi identity in color space. We explicitly check \eqref{eq:id_fa_1} at the end of this appendix. The identity \eqref{eq:id_fa_3} follows from the analogous identity for the 10d gamma matrices
	\begin{equation}
		\Gamma_\mu\=\unity_8\otimes \gamma_\mu\und \Gamma_{3+i}\=2 \begin{pmatrix} 0 & t^i\\ (t^i)^* & 0
		\end{pmatrix} \otimes (\chP^+-\chP^-)\ ,
	\end{equation}
	as well as the anti-commutation relation for the Clebsch-Gordon matrices
	\begin{equation}
		\{t^i,(t^j)^*\}\=-\sfrac12 \delta^{ij}\unity_4\ .
	\end{equation}
	With these identities at hand, we can rewrite the first term in \eqref{eq:ifa_special_case} as
	\begin{equation}
		\begin{aligned}
			-\sfrac12 \mathcal{F}^{\Sigma\Theta}\scrA_\Sigma{\t} \scrA_\Theta
			\ \stackrel{\eqref{eq:id_fa_1}}{=}\ &\sfrac{1}{16} \mathcal{F}^{\Sigma\Theta}\tr\bigl\{(\scrC_{\Sigma})\indices{^4_B}(\scrCb_{\Theta})\indices{^B_C}(\scrC_{\Gamma})\indices{^C_D}(\scrCb_{\Psi})\indices{^D_4}\bigr\} \scrA^\Gamma{\t} \scrA^\Psi\\
			\ \stackrel{\eqref{eq:id_fa_2}}{=}\ &\sfrac{1}{16} \mathcal{F}^{\Sigma\Theta}\tr\bigl\{(\scrC_{\Sigma})\indices{^4_B}(\scrCb_{\Theta})\indices{^B_C}(\scrC_{\Gamma})\indices{^C_D}\scrD^\Gamma\ S\indices{^D_E}\slashed{\scrA}{\indices{^E_F}}\t\slashed{\scrA}^*{\indices{^F_4}}\bigr\}
			\\
			\ \stackrel{\mathrm{ibp}}{=}\ &-\sfrac{1}{16} \scrD^\Gamma \mathcal{F}^{\Sigma\Theta}\tr\bigl\{(\scrC_{\Sigma})\indices{^4_B}(\scrCb_{\Theta})\indices{^B_C}(\scrC_{\Gamma})\indices{^C_D}S\indices{^D_E}\slashed{\scrA}{\indices{^E_F}}\t\slashed{\scrA}^*{\indices{^F_4}}\bigr\}\\
			\ \stackrel{\eqref{eq:id_fa_3}}{=}\ &-\sfrac{1}{16} \scrD^\Gamma \mathcal{F}^{\Sigma\Theta}\tr\bigl\{\bigl[-2(\scrC_{\Sigma})\indices{^4_D}\ \eta_{\Theta\Gamma}+(\scrC_{[\Sigma})\indices{^4_B}(\scrCb_{\Theta})\indices{^B_C}(\scrC_{\Gamma]})\indices{^C_D}\bigr]S\indices{^D_E}\slashed{\scrA}{\indices{^E_F}}\t\slashed{\scrA}^*{\indices{^F_4}}\bigr\}\\
			\=\;&-\sfrac{1}{8} \scrD_\Theta \mathcal{F}^{\Theta\Sigma}\tr\bigl\{(\scrC_{\Sigma})\indices{^4_D}S\indices{^D_E}\slashed{\scrA}{\indices{^E_F}}\t\slashed{\scrA}^*{\indices{^F_4}}\bigr\}\ ,\\
		\end{aligned}
	\end{equation}
	where in the last step we used the Bianchi identity $\scrD^{[\Gamma} \mathcal{F}^{\Sigma\Theta]}=0$. This concludes the proof for the special case
	$L=\operatorname{diag}(-1,-1,-1,+3)$ (and permutations thereof). To reach the full Lie algebra we make use of the fact that we can superimpose coupling flow operators with weight one, giving the Cartan subalgebra and that $S_g^{\mathrm{b}}[\scrA]$ is invariant under R-symmetry transformations $\scrA\rightarrow\scrA'$. From
	\begin{equation}
		0\=(\partial_g+R_g[\scrA'])S_g^{\mathrm{b}}[\scrA']\=(\partial_g+R_g[\scrA'])S_g^{\mathrm{b}}[\scrA]\ ,
	\end{equation}
	we observe the transformed $R_g[\scrA']$ also satisfies the infinitesimal free action condition, reaching all $L\ \in\ \mathfrak{su}(4)$.
	
	Lastly, we prove \eqref{eq:id_fa_1} by explicitly checking the various possibilities of the open indices. The easiest case is the one with only gamma matrices
	\begin{equation}
		\sfrac{1}{4}\tr\bigl\{(\scrC_{\mu})\indices{^4_B}(\scrCb_{\nu})\indices{^B_C}(\scrC_{\rho})\indices{^C_D}(\scrCb_{\sigma})\indices{^D_4}\bigr\}\=\sfrac{1}{4}\tr\bigl\{\gamma_{\mu}\gamma_\nu\gamma_{\rho}\gamma_\sigma\bigr\}\=\eta_{\mu\sigma}\eta_{\nu\rho}-\eta_{\mu\rho}\eta_{\nu\sigma}+\eta_{\mu\nu}\eta_{\rho\sigma}\ .
	\end{equation}
	Next, we consider the case when there are three gamma matrices (modulo chiral projectors) in the trace, i.e.~one of the four indices in the range 4 to 9 and the three others in the range 0 to 3. In that case, the trace vanishes since any trace over an odd number of gamma matrices vanishes and the r.h.s.~of \eqref{eq:id_fa_1} also clearly vanishes because in each term there is a Kronecker delta that is zero. The next case is the one where two indices are in the range 0 to 3 and the other two indices are in the range 4 to 9. We have to distinguish three arrangements of indices
	\begin{equation}
		\begin{aligned}
			\sfrac{1}{4}\tr\bigl\{(\scrC_{3{+}i})\indices{^4_B}(\scrCb_{3{+}j})\indices{^B_C}(\scrC_{\mu})\indices{^C_D}(\scrCb_{\nu})\indices{^D_4}\bigr\}
			&\=\sfrac{1}{4}\tr\bigl\{(\scrC_{3{+}i})\indices{^4_B}(\scrCb_{3{+}j})\indices{^B_4}\gamma_{\mu}\gamma_\nu\bigr\}\\
			&\=(t^{i})_{4J}(t^{j})^{J4}\ \tr\bigl\{\chP^+\gamma_{\mu}\gamma_\nu\bigr\}+(t^{i})^{4J}(t^{j})_{J4}\ \tr\bigl\{\chP^-\gamma_{\mu}\gamma_\nu\bigr\}\\
			&\=-2[(t^{i})_{4J}(t^{j})^{J4}\eta_{\mu\nu}\ +\ \mathrm{c.c.}]
			\=\delta_{ij}\eta_{\mu\nu}
			\ ,
		\end{aligned}
	\end{equation}
	\begin{equation}
		\begin{aligned}
			\sfrac{1}{4}\tr\bigl\{(\scrC_{3{+}i})\indices{^4_B}(\scrCb_{\mu})\indices{^B_C}(\scrC_{3{+}j})\indices{^C_D}(\scrCb_{\nu})\indices{^D_4}\bigr\}
			&\=\tr\bigl\{[(t^{i})_{4J}\chP^+-(t^i)^{4J}\chP^-]\gamma_\mu[(t^{j})_{J4}\chP^+-(t^j)^{J4}\chP^-]\gamma_\nu\bigr\}\\
			&\=-(t^{i})_{4J}(t^j)^{J4}\tr\bigl\{\chP^+\gamma_\mu\gamma_\nu\bigr\}-(t^{i})^{4J}(t^j)_{J4}\tr\bigl\{\chP^-\gamma_\mu\gamma_\nu\bigr\}\\
			&\=2(t^{i})_{4J}(t^j)^{J4}\eta_{\mu\nu}\ +\ \mathrm{c.c.}\=-\delta_{ij}\eta_{\mu\nu}\ ,
		\end{aligned}
	\end{equation}
	\begin{equation}
		\begin{aligned}
			\sfrac{1}{4}\tr\bigl\{(\scrC_{3{+}i})\indices{^4_B}(\scrCb_{\mu})\indices{^B_C}(\scrC_{\nu})\indices{^C_D}(\scrCb_{3{+}j})\indices{^D_4}\bigr\}
			&\=\tr\bigl\{[(t^{i})_{4J}\chP^+-(t^i)^{4J}\chP^-]\gamma_\mu\gamma_\nu[(t^{j})^ {J4}\chP^+-(t^j)_{J4}\chP^-]\bigr\}\\
			&\=(t^{i})_{4J}(t^{j})^{J4}\ \tr\bigl\{\chP^+\gamma_{\mu}\gamma_\nu\bigr\}+(t^{i})^{4J}(t^{j})_{J4}\ \tr\bigl\{\chP^-\gamma_{\mu}\gamma_\nu\bigr\}\\
			&\=-2[(t^{i})_{4J}(t^{j})^{J4}\eta_{\mu\nu}\ +\ \mathrm{c.c.}]
			\=\delta_{ij}\eta_{\mu\nu}
			\ ,
		\end{aligned}
	\end{equation}
	with all the other index configurations related to the three above by the cyclicity of the trace. The trace with only one gamma matrix vanishes due to the same reason as for three gamma matrices. We are left with the case
	\begin{equation}
		\begin{aligned}
			\sfrac{1}{4}\tr\bigl\{(\scrC_{3{+}i})\indices{^4_B}(\scrCb_{3{+}j})\indices{^B_C}(\scrC_{3{+}k})\indices{^C_D}(\scrCb_{3{+}l})\indices{^D_4}\bigr\}
			\=&4(t^i){\indices{_{4I}}}(t^j){\indices{^{IC}}}(t^k){\indices{_{CK}}}(t^l){\indices{^{K4}}}\ \tr\ \chP^+\\
			+&4(t^i){\indices{^{4I}}}(t^j){\indices{_{IC}}}(t^k){\indices{^{CK}}}(t^l){\indices{_{K4}}}\ \tr\ \chP^-\\
			\=&8\ (t^i){\indices{_{4I}}}(t^j){\indices{^{IC}}}(t^k){\indices{_{CK}}}(t^l){\indices{^{K4}}}\ +\ \mathrm{c.c.}\\
			\=&8\ [(t^i){\indices{_{4I}}}(t^j){\indices{^{I4}}}(t^k){\indices{_{4K}}}(t^l){\indices{^{K4}}}\\
			&+(t^i){\indices{_{4I}}}(t^j){\indices{^{IJ}}}(t^k){\indices{_{JK}}}(t^l){\indices{^{K4}}}]\ +\ \mathrm{c.c.}
		\end{aligned}
	\end{equation}
	The last expression can be evaluated with the explicit form of the Clebsch-Gordon coefficients \eqref{eq:CG_coeff}
	and the identity
	\begin{equation}
		\epsilon_{IJM}\epsilon^{MKL}\=\delta\indices{_I^K}\delta\indices{_J^L}-\delta\indices{_I^L}\delta\indices{_J^K}\ .
	\end{equation}
	We do not quite find the desired result, because we obtain additional terms when two of the indices $i,j,k,l$ are in the range 1 to 3 and the other two are in the range 4 to 6. For example
	\begin{equation}
		\sfrac{1}{4}\tr\bigl\{(\scrC_{3{+}I})\indices{^4_B}(\scrCb_{6{+}J})\indices{^B_C}(\scrC_{3{+}K})\indices{^C_D}(\scrCb_{6{+}L})\indices{^D_4}\bigr\}\=\delta_{IL}\delta_{JK}-\delta_{IK}\delta_{JL}-\delta_{IJ}\delta_{KL}\ ,
	\end{equation}
	where only the second term on the r.h.s.~would appear in the r.h.s.~of \eqref{eq:id_fa_1}.
	However, we contract \eqref{eq:id_fa_1} with the $\varphi$'s in the adjoint representation of the gauge group. It turns out that the additional terms are proportional to
	\begin{equation}
		(\varphi_{I}{\t}\varphi_{J})\ (\varphi_{I+3}{\t}\varphi_{J+3})
		+(\varphi_{I}{\t}\varphi_{J+3})\ (\varphi_{J}{\t}\varphi_{I+3})+
		(\varphi_{I}{\t}\varphi_{I+3})\ (\varphi_{J+3}{\t}\varphi_{J})\=0\ ,
	\end{equation}
	i.e.~vanish by the Jacobi identity in color space.
	
	\section{Explicit computation of the Nicolai maps}\label{app:formulae}
	We first collect a number of useful identities
	\begin{equation}
		\begin{aligned}
			&\sum_B \Phi\indices{^A_B}\t(\Phi\indices{^B_A})^{\+}\=\left\{\begin{array}{cl}
				-2\gamma_5\sum_J\varphi_J\t\varphi_{J+3} & \text{for}\quad A\=4\\
				2\gamma_5\sum_J(-)^{\delta_{KJ}}\varphi_J\t\varphi_{J+3}  & \text{for}\quad A\=K\\
			\end{array}\right.\ ,\\
			&\sum_B \Phi\indices{^A_B}S_0\Phi\indices{^B_A}\=\left\{\begin{array}{cl}
				\sum_j\varphi_jS_0\varphi_j-\gamma_5\sum_J(\varphi_{J+3}S_0\varphi_J-\varphi_JS_0\varphi_{J+3}) & \text{for}\quad A\=4\\
				\sum_j\varphi_jS_0\varphi_j+\gamma_5\sum_J(-)^{\delta_{KJ}}(\varphi_{J+3}S_0\varphi_J-\varphi_JS_0\varphi_{J+3}) & \text{for}\quad A\=K\\
			\end{array}\right.\ ,\\
			&\sum_B (c^{I})\indices{^A_B}S_0\Phi\indices{^B_A}\=\left\{\begin{array}{cl}
				S_0\varphi_I+\gamma_5 S_0\varphi_{I+3} & \text{for}\quad A\=4\\
				S_0\varphi_I-\gamma_5 S_0\varphi_{I+3}(-)^{\delta_{IK}}  &\text{for}\quad A\=K\\
			\end{array}\right.\ ,\\
			&\sum_B (c^{I{+}3})\indices{^A_B}S_0\Phi\indices{^B_A}\=\left\{\begin{array}{cl}
				-\gamma_5 S_0\varphi_I+ S_0\varphi_{I+3} & \text{for}\quad A\=4\\
				\gamma_5 S_0\varphi_I(-)^{\delta_{IK}}+ S_0\varphi_{I+3}  &\text{for}\quad A\=K\\
			\end{array}\right.\ ,
		\end{aligned}
	\end{equation}
	and
	\begin{equation}\label{eq:tr4}
		\begin{aligned}
			\sum_{B,C,D}\tr\bigl\{(c^{I})\indices{^4_B}S_0 \Phi\indices{^B_C}S_0 \Phi\indices{^C_D}\t(\Phi\indices{^D_4})^{\+}\bigr\}&\=\tr\Bigl\{ 
			-2S_0\sum_j\varphi_jS_0 \varphi_j \varphi_I\\
			-2S_0\sum_J&\Bigl[\varphi_{I+3}S_0\varphi_J\varphi_{J+3}
			-\varphi_{J}S_0\varphi_{I+3}\varphi_{J+3}+\varphi_{J+3}S_0\varphi_{I+3}\varphi_{J}\Bigr]\Bigr\}\ ,
		\end{aligned}
	\end{equation}
	\begin{equation}\label{eq:trK}
		\begin{aligned}
			\sum_{B,C,D}\tr\bigl\{(c^{I})\indices{^K_B}S_0 \Phi\indices{^B_C}S_0 \Phi\indices{^C_D}\t(\Phi\indices{^D_K})^{\+}\bigr\}&\=\tr\Bigl\{
			-2S_0\sum_j\varphi_jS_0 \varphi_j \varphi_I\\
			-2S_0\sum_J&\Bigl[\varphi_{I+3}S_0\varphi_J\varphi_{J+3}
			-\varphi_{J}S_0\varphi_{I+3}\varphi_{J+3}+\varphi_{J+3}S_0\varphi_{I+3}\varphi_{J}\Bigr](-)^{\delta_{IK}}\\
			+4S_0&\Bigl[\varphi_{I+3}S_0\varphi_K\varphi_{K+3}
			-\varphi_{K}S_0\varphi_{I+3}\varphi_{K+3}+\varphi_{K+3}S_0\varphi_{I+3}\varphi_{K}\Bigr]\Bigr\}\ ,
		\end{aligned}
	\end{equation}
	where in \eqref{eq:tr4} and \eqref{eq:trK} half of the terms dropped out due to $\tr\ \gamma^\mu\gamma^\nu\gamma_5=0$. The analogous formulae with $(c^{I{+}3})\indices{^A_B}$ are given by \eqref{eq:tr4} and \eqref{eq:trK} with all indices $I,J,K,...$ replaced by $I{+}3,J{+}3,K{+}3,...$ and vice versa so that e.g.~
	\begin{equation}
		\begin{aligned}
			\sum_{B,C,D}\tr\bigl\{(c^{I{+}3})\indices{^4_B}S_0 \Phi\indices{^B_C}S_0 \Phi\indices{^C_D}\t(\Phi\indices{^D_4})^{\+}\bigr\}&\=\tr\Bigl\{ 
			-2S_0\sum_j\varphi_jS_0 \varphi_j \varphi_{I+3}\\
			-2S_0\sum_{J}&\Bigl[\varphi_{I}S_0\varphi_{J+3}\varphi_{J}
			-\varphi_{J+3}S_0\varphi_{I}\varphi_{J}+\varphi_{J}S_0\varphi_{I}\varphi_{J+3}\Bigr]\Bigr\}\ .
		\end{aligned}
	\end{equation}
	To evaluate the traces, we need
	\begin{equation}
		\begin{aligned}
			\tr\ \gamma_5 \gamma^\mu\gamma^\nu\gamma^\rho\gamma^\sigma&\=-4\epsilon^{\mu\nu\rho\sigma}\ ,\\
			\tr\ \gamma^\mu\gamma^\nu\gamma^\rho\gamma^\sigma&\=4(\eta^{\mu\nu}\eta^{\rho\sigma}-\eta^{\mu\rho}\eta^{\nu\sigma}+\eta^{\mu\sigma}\eta^{\nu\rho})\ ,\\
			\tr\ \gamma^\mu\gamma^\nu\gamma^\rho\gamma^\sigma\gamma^\lambda\gamma^\eta&\=-\eta^{\mu\nu}\ \tr\ \gamma^\rho\gamma^\sigma\gamma^\lambda\gamma^\eta+\eta^{\mu\rho}\ \tr\ \gamma^\nu\gamma^\sigma\gamma^\lambda\gamma^\eta\mp...\ .
		\end{aligned}
	\end{equation}
	We now give some intermediate results for the various contributions to the Nicolai maps.
	
	\noindent\textbf{First order:}
	We have
	\begin{equation}
		\stackrel{\leftarrow}{R_1}{^{(A)}}\=\sfrac18\stackrel{\longleftarrow}{\sfrac{\delta}{\delta\scrA_\Gamma}}\Pi\indices{_\Gamma^\Sigma}\ \tr\Bigl\{(\scrC_{\Sigma})\indices{^A_B}S_0\slashed{\scrA}{\indices{^B_C}}\times \slashed{\scrA}^*{\indices{^{C}_A}}\Bigl\}\ ,
	\end{equation}
	where $\Pi\indices{_\Gamma^\Sigma}\=\delta\indices{_\Gamma^\Sigma}-C\indices{_\Gamma^\Sigma}$ and $\partial_{3+i} \equiv 0$. Further expanding
	\begin{equation}
		\stackrel{\leftarrow}{R_1}{^{(A)}}\=\sfrac18\stackrel{\longleftarrow}{\sfrac{\delta}{\delta\scrA_\Gamma}}\Pi\indices{_\Gamma^\Sigma}\ \tr\Bigl\{(\scrC_{\Sigma})\indices{^A_B}S_0\Bigl[\slA\t\slA\delta\indices{^B_A}+2\Phi\indices{^B_A}\t\slA+\Phi\indices{^B_C}\t(\Phi\indices{^C_A})^{\+}\Bigr]\Bigl\}\ ,
	\end{equation}
	we find 
	\begin{equation}
		R_1{^{(A)}} A_\mu \=\sfrac{1}{8}\Pi\indices{_\mu^\nu}\ \tr\Bigl\{\gamma_\nu S_0 \slA\t\slA\Bigr\}\=C^\rho A_\mu A_\rho\ ,\quad \forall\ A\=1,2,3,4\ ,
	\end{equation}
	where we used $(\scrC_{\nu})\indices{^A_J}\=0$, as well as $\Phi\indices{^A_J}\t(\Phi\indices{^J_A})^{\+}\ \sim\ \gamma_5 $ and $\tr\ {\gamma_\nu\gamma_\rho \gamma_5}\=0$. The remaining part is
	\begin{equation}
		R_1{^{(A)}} \varphi_i \=\sfrac{1}{4}\ \tr\Bigl\{(c^{i})\indices{^A_B} S_0 \Phi\indices{^B_A}\t\slA\Bigr\}\=C^\rho \varphi_iA_\rho\ ,\quad \forall\ A\=1,2,3,4\ ,
	\end{equation}
	where we used that a trace over an odd number of gamma matrices vanishes.
	
	\noindent\textbf{Second order:}
	First, it is straightforward to obtain
	\begin{equation}
		(R_1{^{(A)}})^2 A_\mu\=2C^\rho A_{[\mu}C^\lambda A_{\rho]}A_\lambda\ ,\quad \forall\ A\=1,2,3,4\ ,
	\end{equation}
	\begin{equation}
		(R_1{^{(A)}})^2 \varphi_i\=C_\rho \varphi_i C_\lambda A^\rho A^\lambda-C_\rho A^\rho C_\lambda \varphi_i A^\lambda\ \quad \forall\ A\=1,2,3,4\ .
	\end{equation}
	With the perturbative expansion of the covariant projector 
	\begin{equation}
		P\indices{_\Gamma^\Sigma}\=\delta\indices{_\Gamma^\Sigma}-\scrD_\Gamma G \partial^\Sigma
		\=\Pi\indices{_\Gamma^\Theta}\bigl\{\delta\indices{_\Theta^\Sigma}-g\scrA_\Theta\sum_{k=0}^{\infty}(-g\partial{\cdot}A\;\; C)^kC\partial^\Sigma\bigr\}\ ,
	\end{equation}
	and the gaugino propagator \eqref{eq:perturb_S}, we further find
	\begin{equation}
		\stackrel{\leftarrow}{R_2}{^{(A)}}\=-\sfrac18\stackrel{\longleftarrow}{\sfrac{\delta}{\delta\scrA_\Gamma}}\Pi\indices{_\Gamma^\Sigma}\ \tr\Bigl\{(\scrC_{\Sigma}){\indices{^A_B}}S_0\slashed{\scrA}{\indices{^B_C}}S_0\slashed{\scrA}{\indices{^C_D}}\t \slashed{\scrA}^*{\indices{^D_A}}\Bigl\}
		-\sfrac18\stackrel{\longleftarrow}{\sfrac{\delta}{\delta\scrA_\Gamma}}\Pi\indices{_\Gamma^\Theta}\scrA_\Theta S_0\partial^\sigma\tr\Bigl\{\gamma_\sigma S_0\slashed{\scrA}{\indices{^A_B}}\t \slashed{\scrA}^*{\indices{^B_A}}\Bigl\}\ .
	\end{equation}
	It is easy to see that the second term gives no contribution using $\Phi{\indices{^A_B}}\t(\Phi{\indices{^B_A}})^{\+}\ \sim\ \gamma_5 $ and symmetry. Evaluating the traces, one finds
	\begin{equation}
		\begin{aligned}
			R_2{^{(4)}} A_\mu
			&\=3C^\rho A^\lambda C_{[\mu}A_\lambda A_{\rho]}+2C^\rho A_{[\mu} C^\lambda  A_{\rho]} A_\lambda+2C^\rho \varphi_i C_{[\rho}  A_{\mu]}\varphi_i\\
			&\qquad+\Pi\indices{_\mu^\nu}\epsilon_{\nu\lambda\rho\sigma}\sum_{J=1}^{3}[C^\lambda\varphi_J C^\rho \varphi_{J+3}A^\sigma-C^\lambda\varphi_{J+3} C^\rho \varphi_{J}A^\sigma +C^\lambda A^\rho C^\sigma\varphi_{J+3}\varphi_J]\ ,
		\end{aligned}
	\end{equation}
	\begin{equation}
		\begin{aligned}
			R_2{^{(K)}} A_\mu
			&\=3C^\rho A^\lambda C_{[\mu}A_\lambda A_{\rho]}+2C^\rho A_{[\mu} C^\lambda  A_{\rho]} A_\lambda+2C^\rho \varphi_i C_{[\rho}  A_{\mu]}\varphi_i\\
			&\qquad-\Pi\indices{_\mu^\nu}\epsilon_{\nu\lambda\rho\sigma}\sum_{J=1}^{3}(-)^{\delta_{KJ}}[C^\lambda\varphi_J C^\rho \varphi_{J+3}A^\sigma-C^\lambda\varphi_{J+3} C^\rho \varphi_{J}A^\sigma +C^\lambda A^\rho C^\sigma\varphi_{J+3}\varphi_J]\ ,
		\end{aligned}
	\end{equation}
	and
	\begin{equation}\label{eq:temp}
		\begin{aligned}
			R_2{^{(4)}}\varphi_I\=& C^\rho \varphi_I C^\lambda A_\rho A_\lambda-C_\rho A^\rho C_\lambda \varphi_I A^\lambda+2C^{[\rho}A^{\lambda]}C_\rho \varphi_IA_\lambda+C^\rho \varphi_{j}C_\rho \varphi_I\varphi_j\\+&\sfrac12 \epsilon_{\mu\nu\rho\lambda}[C^\mu \varphi_{I+3} C^\nu A^\rho A^\lambda + 2C^\mu A^\nu C^\rho \varphi_{I+3}A^\lambda]\\
			+&C_\rho\sum_{J=1}^{3}\bigl[\varphi_{I+3}C_\rho \varphi_{J+3}\varphi_J+\varphi_{J}C_\rho \varphi_{I+3}\varphi_{J+3}-\varphi_{J+3} \varphi_{I+3}\varphi_{J}\bigr]\ ,
		\end{aligned}
	\end{equation}
	\begin{equation}
		\begin{aligned}
			R_2{^{(K)}}\varphi_I\=& C^\rho \varphi_I C^\lambda A_\rho A_\lambda-C_\rho A^\rho C_\lambda \varphi_I A^\lambda+2C^{[\rho}A^{\lambda]}C_\rho 	\varphi_IA_\lambda+C^\rho \varphi_{j}C_\rho \varphi_I\varphi_j\\-&\sfrac12 \epsilon_{\mu\nu\rho\lambda}(-)^{\delta_{IK}}[C^\mu \varphi_{I+3} C^\nu A^\rho A^\lambda + 2C^\mu A^\nu C^\rho \varphi_{I+3}A^\lambda]\\
			+&C^\rho(-)^{\delta_{IK}}\sum_{J=1}^{3}\bigl[ \varphi_{I+3}C_\rho \varphi_{J+3}\varphi_J+ \varphi_{J}C_\rho \varphi_{I+3}\varphi_{J+3}- 	\varphi_{J+3}C_\rho \varphi_{I+3}\varphi_{J}\bigr]\\
			-&2C^\rho\bigl[ \varphi_{I+3}C_\rho \varphi_{K+3}\varphi_K+\varphi_{K}C_\rho \varphi_{I+3}\varphi_{K+3}-\varphi_{K+3}C_\rho 	\varphi_{I+3}\varphi_{K}\bigr]\ ,
		\end{aligned}
	\end{equation}
	\begin{equation}
		\begin{aligned}
			R_2{^{(4)}}\varphi_{I+3}\=& C^\rho \varphi_{I+3} C^\lambda A_\rho A_\lambda-C_\rho A^\rho C_\lambda \varphi_{I+3} A^\lambda+2C^{[\rho}A^{\lambda]}C_\rho \varphi_{I+3}A_\lambda+C^\rho \varphi_{j}C_\rho \varphi_{I+3}\varphi_j\\-&\sfrac12 \epsilon_{\mu\nu\rho\lambda}[C^\mu \varphi_{I} C^\nu A^\rho A^\lambda + 2C^\mu A^\nu C^\rho \varphi_{I}A^\lambda]\\
			-&C_\rho\sum_{J=1}^{3}\bigl[\varphi_{I}C_\rho \varphi_{J+3}\varphi_J+\varphi_{J}C_\rho \varphi_{I}\varphi_{J+3}-\varphi_{J+3} \varphi_{I}\varphi_{J}\bigr]\ ,
		\end{aligned}
	\end{equation}
	\begin{equation}
		\begin{aligned}
			R_2{^{(K)}}\varphi_{I+3}\=& C^\rho \varphi_{I+3} C^\lambda A_\rho A_\lambda-C_\rho A^\rho C_\lambda \varphi_{I+3} A^\lambda+2C^{[\rho}A^{\lambda]}C_\rho \varphi_{I+3}A_\lambda+C^\rho \varphi_{j}C_\rho \varphi_{I+3}\varphi_j\\+&\sfrac12 \epsilon_{\mu\nu\rho\lambda}(-)^{\delta_{IK}}[C^\mu \varphi_{I} C^\nu A^\rho A^\lambda + 2C^\mu A^\nu C^\rho \varphi_{I}A^\lambda]\\
			-&C_\rho(-)^{\delta_{IK}}\sum_{J=1}^{3}\bigl[ \varphi_{I}C_\rho \varphi_{J+3}\varphi_J+ \varphi_{J}C_\rho \varphi_{I}\varphi_{J+3}-C^\rho \varphi_{J+3} \varphi_{I}\varphi_{J}\bigr]\\
			+&2\bigl[ \varphi_{I}C_\rho \varphi_{K+3}\varphi_K+ \varphi_{K}C_\rho \varphi_{I}\varphi_{K+3}-C^\rho \varphi_{K+3} \varphi_{I}\varphi_{K}\bigr]\ ,
		\end{aligned}
	\end{equation}
	for $K=1,2,3$.

	\section{Testing the conditions of the Nicolai maps}\label{app:tests}
	We explicitly check the three conditions for the four distinct Nicolai maps \eqref{eq:NM_dr1_4}-\eqref{eq:NM_dr2_K}.
	
	\noindent\textbf{Check of the gauge condition:} The (Landau) gauge condition $\partial^\mu T_g A_\mu=\partial^\mu A_\mu+\mathcal{O}(g^3)$ follows from symmetry and $\partial^\mu\Pi\indices{_\mu^\nu}=0$.
	
	\noindent\textbf{Check of the free-action condition:} The maps have to satisfy
	\begin{equation}
		S_0[A'_\mu, \varphi'_I, \varphi'_{I+3}]\=S^{\mathrm{b}}_g[A_\mu, \varphi_I, \varphi_{I+3}]\ ,
	\end{equation}
	with the bosonic action
	\begin{equation}
		\begin{aligned}
			&S^{\mathrm{b}}_g\=\intdx \bigl\{-\sfrac{1}{4}F_{\mu\nu}F^{\mu\nu}-\sfrac12 \Diff_\mu \varphi_i \Diff^\mu \varphi_i -\sfrac{g^2}{4} (\varphi_i\t\varphi_j)^2 \bigr\}\ ,\\
			&F_{\mu\nu}\=\partial_\mu A_\nu-\partial_\nu A_\mu +gA_\mu\t A_\nu\ ,\\
			&\Diff_\mu \=\partial_\mu + gA_\mu \t \ \ ,
		\end{aligned}
	\end{equation}
	and $S^{\mathrm{b}}_0=S^{\mathrm{b}}_{g=0}$. Writing this condition out at second order gives
	\begin{equation}\label{eq:fa_cond}
		\begin{aligned}
			\intdx \Bigl\{&\sfrac12 A'_\mu|_{\mathcal{O}(g)}\bigl(\Box\eta^{\mu\nu}-\partial^\mu\partial^\nu \bigr)A'_\nu|_{\mathcal{O}(g)}+ A_\mu\bigl(\Box\eta^{\mu\nu}-\partial^\mu\partial^\nu \bigr)A'_\nu|_{\mathcal{O}(g^2)}+\sfrac12\varphi'_i|_{\mathcal{O}(g)}\Box\varphi'_i|_{\mathcal{O}(g)}+\varphi_i\Box\varphi'_i|_{\mathcal{O}(g^2)}\Bigr\}\\
			&\=\intdx\Bigl\{-\sfrac14(A_\mu\t A_\nu)^2-\sfrac12(A_\mu\t\varphi_i)^2-\sfrac14(\varphi_i\t\varphi_j)^2\Bigr\}\ ,
		\end{aligned}
	\end{equation}
	after integrating by parts on the left hand side.
	The free-action condition (in the Landau gauge) was previously shown for the map in $\mathcal{N}=\,1$ $D=10$ SYM, i.e.~before dimensional reduction. It is clear that the condition remains valid for the reduced map (given by the black terms) and action. We therefore argue that we only have left to show that the blue terms have no effect on the l.h.s.~of \eqref{eq:fa_cond}. Further using $\partial^\mu A'_\mu=0$ (at all orders), we can drop two terms on the l.h.s.~of \eqref{eq:fa_cond}, so we are left with the condition
	\begin{equation}
		\intdx\Bigl\{A_\mu\Box A'^\mu|_{\mathcal{O}(g^2)}+\varphi_I\Box \varphi'_I|_{\mathcal{O}(g^2)}+\varphi_{I+3}\Box \varphi'_{I+3}|_{\mathcal{O}(g^2)}\Bigr\}_{\textcolor{blue}{\text{blue terms}}}\=0\ .
	\end{equation}
	Let us start with the map obtained from $A=4$.
	We refer to the three contributions as $\circled{1}$, $\circled{2}$, $\circled{3}$ respectively. Dividing by an overall factor of $\sfrac12$ and switching to a graphical notation, we find for the first part
	\begin{equation}
		\circled{1}\=\epsilon_{\mu\nu\rho\lambda}\Biggl\{
		\sdiagram{\partial^\nu A^\mu}{\varphi_J}{C^\rho}{\varphi_{J+3}}{A^\lambda}-
		\sdiagram{\partial^\nu A^\mu}{\varphi_{J+3}}{C^\rho}{\varphi_{J}}{A^\lambda}+
		\sdiagram{\partial^\nu A^\mu}{A^\rho}{C^\lambda}{\varphi_{J+3}}{\varphi_{J}}\Biggr\}\ .
	\end{equation}
	The last diagram drops out since we can bring the $\partial^\nu$ to the center through integration by parts and the overall anti-symmetry under $\mu\leftrightarrow\rho$ (gaining a factor $-1/2$). Then, clearly $\partial^\nu C^\lambda=C^{\nu\lambda}$ contracts to zero with the epsilon symbol as it is symmetric under $\nu\leftrightarrow\lambda$.
	The second contribution is
	\begin{equation}\label{eq:contr2}
		\begin{aligned}
			\circled{2}\=&\sfrac12 \epsilon_{\mu\nu\rho\lambda} \Biggl\{
			\sdiagram{\partial^\mu \varphi_I}{\varphi_{I+3}}{C^\nu}{A^\rho}{A^\lambda}+2
			\sdiagram{\partial^\mu \varphi_I}{A^\nu}{C^\rho}{\varphi_{I+3}}{A^\lambda}\Biggr\}\\
			&+\Biggl\{
			\sdiagram{\partial_\rho \varphi_I}{\varphi_{I+3}}{C^\rho}{\varphi_{J+3}}{\varphi_J}+
			\sdiagram{\partial_\rho \varphi_I}{\varphi_{J}}{C^\rho}{\varphi_{I+3}}{\varphi_{J+3}}-
			\sdiagram{\partial_\rho \varphi_I}{\varphi_{J+3}}{C^\rho}{\varphi_{I+3}}{\varphi_J}
			\Biggr\}
		\end{aligned}
	\end{equation}
	and
	\begin{equation}\label{eq:contr3}
		\circled{3}\=-\bigl[\circled{2}\ \text{with}\ (I\leftrightarrow I{+}3)\bigr]\ .
	\end{equation}
	The first respective terms of $\circled{2}$ and $\circled{3}$ cancel each other by means of integration by parts and symmetry. The second term of $\circled{2}$ cancels the first term of $\circled{1}$ and the corresponding second term of $\circled{3}$ cancels the second term of $\circled{1}$. We are left with the second line of \eqref{eq:contr2} and the corresponding terms from \eqref{eq:contr3}:
	\begin{equation}\label{eq:six_terms}
		\begin{aligned}
			&
			\sdiagram{\partial_\rho \varphi_I}{\varphi_{I+3}}{C^\rho}{\varphi_{J+3}}{\varphi_J}+
			\sdiagram{\partial_\rho \varphi_I}{\varphi_{J}}{C^\rho}{\varphi_{I+3}}{\varphi_{J+3}}-
			\sdiagram{\partial_\rho \varphi_I}{\varphi_{J+3}}{C^\rho}{\varphi_{I+3}}{\varphi_J}\\
			-&
			\sdiagram{\partial_\rho \varphi_{I+3}}{\varphi_{I}}{C^\rho}{\varphi_{J+3}}{\varphi_J}-
			\sdiagram{\partial_\rho \varphi_{I+3}}{\varphi_{J}}{C^\rho}{\varphi_{I}}{\varphi_{J+3}}+
			\sdiagram{\partial_\rho \varphi_{I+3}}{\varphi_{J+3}}{C^\rho}{\varphi_{I}}{\varphi_J}\ =:\ \sum_{I,J}Z_{IJ}\ \stackrel{!}{=}\ 0\ .
		\end{aligned}
	\end{equation}
	We can integrate by parts in the first diagram, which gives two contributions, one of them canceling the fourth diagram. Similarly, the third and fifth diagram can be combined into one. In the second and last diagram we can make use of the anti-symmetry under $I\leftrightarrow J$ to also integrate by parts in both diagrams (gaining a factor $-1/2$ in each) and then combine them into one contribution. This way the six diagrams reduce to three:
	\begin{equation}\label{eq:fa_jacobi}
		-\xdiagram{\varphi_I}{\varphi_{I+3}}{\varphi_{J+3}}{\varphi_J}-\xdiagram{\varphi_I}{\varphi_{J+3}}{\varphi_{J}}{\varphi_{I+3}}-\xdiagram{\varphi_I}{\varphi_{J}}{\varphi_{I+3}}{\varphi_{J+3}}\ \stackrel{!}{=}\ 0\ ,
	\end{equation}
	where we used $\partial_\rho C^\rho = \Box C = \unity$. The condition \eqref{eq:fa_jacobi} follows simply from the Jacobi identity (in color space). In conclusion, the blue terms of the $A=4$ map indeed have no effect on the free action condition, at least to the second order.
	
	\noindent For the $A=K$ case, most of the calculation can be done in the same way as for $A=4$ by simply carrying around the sign factors $(-)^{\delta_{KJ}}$ etc. However, the third and last line in \eqref{eq:NM_dr2_K} require special attention. Referring to \eqref{eq:six_terms}, the remaining condition for $A=K$ can be written as
	\begin{equation}
		\sum_{I,J}(-)^{\delta_{KI}}Z_{IJ}-2\sum_I Z_{IK}\=\sum_J\bigl(-Z_{KJ}-2Z_{JK}+\sum_{I\neq K}Z_{IJ}\bigr)\ \stackrel{!}{=}\ 0\ .
	\end{equation}
	In the following we will show that this condition is satisfied by demonstrating that 
	\begin{equation}
		\sum_JZ_{JK}\=-\sum_JZ_{KJ}\ ,\quad \text{ for any }\quad K\=1,2,3\ ,
	\end{equation}
	and using our previous result $\sum_{I,J} Z_{IJ}=0$. In order to make clear that we are not summing over $K$, we will set $K=1$ in the following, although the calculation works in the exact same way for $K=2,3$. We start with
	\begin{equation}
		\begin{aligned}
			\sum_JZ_{J1}\=&
			\sdiagram{\partial_\rho \varphi_J}{\varphi_{J+3}}{C^\rho}{\varphi_{4}}{\varphi_1}+
			\sdiagram{\partial_\rho \varphi_J}{\varphi_{1}}{C^\rho}{\varphi_{J+3}}{\varphi_{4}}-
			\sdiagram{\partial_\rho \varphi_J}{\varphi_{4}}{C^\rho}{\varphi_{J+3}}{\varphi_1}\\
			-&
			\sdiagram{\partial_\rho \varphi_{J+3}}{\varphi_{J}}{C^\rho}{\varphi_{4}}{\varphi_1}-
			\sdiagram{\partial_\rho \varphi_{J+3}}{\varphi_{1}}{C^\rho}{\varphi_{J}}{\varphi_{4}}+
			\sdiagram{\partial_\rho \varphi_{J+3}}{\varphi_{4}}{C^\rho}{\varphi_{J}}{\varphi_1}\ .
		\end{aligned}
	\end{equation}
	We can integrate by parts in the first diagram as previously. This gives two diagrams of which one cancels with the fourth diagram. Further, we integrate by parts in all of the four other diagrams. This leaves us with the 9 diagrams
	\begin{equation}
		\begin{aligned}
			\sum_JZ_{J1}\=
			&
			-\xdiagram{\varphi_J}{\varphi_{J+3}}{\varphi_{4}}{\varphi_1}
			-\xdiagram{\varphi_J}{\varphi_{1}}{\varphi_{J+3}}{\varphi_{4}}
			-\xdiagram{\varphi_J}{\varphi_{4}}{\varphi_{1}}{\varphi_{J+3}}\\
			&
			+\xdiagram{\varphi_{J+3}}{\varphi_{1}}{\varphi_{J}}{\varphi_4}
			-\xdiagram{\varphi_{J+3}}{\varphi_{4}}{\varphi_{J}}{\varphi_{1}}
			-\sdiagram{ \varphi_J}{\partial_\rho \varphi_{1}}{C^\rho}{\varphi_{J+3}}{\varphi_4}\\
			&
			+\sdiagram{\varphi_{J}}{\partial_\rho\varphi_{4}}{C^\rho}{\varphi_{J+3}}{\varphi_1}
			+\sdiagram{\varphi_{J+3}}{\partial_\rho\varphi_{1}}{C^\rho}{\varphi_{J}}{\varphi_{4}}
			-\sdiagram{\varphi_{J+3}}{\partial_\rho\varphi_{4}}{C^\rho}{\varphi_{J}}{\varphi_1}\ ,
		\end{aligned}
	\end{equation}
	of which the first three cancel through the Jacobi identity and the fourth and fifth can be combined into one using the Jacobi identity once more. We find
	\begin{equation}
		\begin{aligned}
			\sum_JZ_{J1}\=
			&
			+\xdiagram{\varphi_{J+3}}{\varphi_{J}}{\varphi_{1}}{\varphi_4}
			-\sdiagram{ \varphi_J}{\partial_\rho \varphi_{1}}{C^\rho}{\varphi_{J+3}}{\varphi_4}\\
			&
			+\sdiagram{\varphi_{J}}{\partial_\rho\varphi_{4}}{C^\rho}{\varphi_{J+3}}{\varphi_1}
			+\sdiagram{\varphi_{J+3}}{\partial_\rho\varphi_{1}}{C^\rho}{\varphi_{J}}{\varphi_{4}}
			-\sdiagram{\varphi_{J+3}}{\partial_\rho\varphi_{4}}{C^\rho}{\varphi_{J}}{\varphi_1}\ .
		\end{aligned}
	\end{equation}
	We can use $\unity=C\indices{^\rho_\rho}$, symmetry and integration by parts to modify the first diagram
	\begin{equation}
		\begin{aligned}
			\xdiagram{\varphi_{J+3}}{\varphi_{J}}{\varphi_{1}}{\varphi_4}\=\xdiagram{\varphi_{1}}{\varphi_4}{\varphi_{J+3}}{\varphi_{J}}\=&\sdiagram{\varphi_{1}}{\varphi_4}{\partial_\rho C^\rho}{\varphi_{J+3}}{\varphi_{J}}\\
			\=&-\sdiagram{\partial_\rho\varphi_{1}}{\varphi_4}{ C^\rho}{\varphi_{J+3}}{\varphi_{J}}-\sdiagram{\varphi_{1}}{\partial_\rho\varphi_4}{ C^\rho}{\varphi_{J+3}}{\varphi_{J}}
		\end{aligned}
	\end{equation}
	Using antisymmetry to flip the external lines appropriately, it is now easy to see that
	\begin{equation}
		\begin{aligned}
			\sum_JZ_{J1}\=
			&
			-\sdiagram{\partial_\rho\varphi_{1}}{\varphi_4}{ C^\rho}{\varphi_{J+3}}{\varphi_{J}}
			+\sdiagram{\partial_\rho\varphi_4}{\varphi_{1}}{ C^\rho}{\varphi_{J+3}}{\varphi_{J}}
			-\sdiagram{\partial_\rho \varphi_{1}}{ \varphi_J}{C^\rho}{\varphi_4}{\varphi_{J+3}}\\
			&
			+\sdiagram{\partial_\rho\varphi_{4}}{\varphi_{J}}{C^\rho}{\varphi_1}{\varphi_{J+3}}
			+\sdiagram{\partial_\rho\varphi_{1}}{\varphi_{J+3}}{C^\rho}{\varphi_{4}}{\varphi_{J}}
			-\sdiagram{\partial_\rho\varphi_{4}}{\varphi_{J+3}}{C^\rho}{\varphi_1}{\varphi_{J}}\=-\sum_JZ_{1J}\ ,
		\end{aligned}
	\end{equation}
	which concludes our check of the free action condition for $A=K$.

	\noindent\textbf{Check of the determinant matching:} The map has to satisfy
	\begin{equation}\label{eq:det_mat}
		\operatorname{log}\ \operatorname{det}\ \frac{\delta \scrA'}{\delta \scrA}\=\operatorname{log}\ \MSS\FP\ ,
	\end{equation}
	which has been checked for the $\mathcal{N}=\,1$ $D=10$ result (to fourth order). It is easy to convince oneself that the condition is preserved under dimensional reduction to $\mathcal{N}=\,4$ $D=4$. Hence, we will again only show that the blue terms have no effect on the l.h.s.~of the condition. Using $\operatorname{log}\ \operatorname{det}=\tr\ \operatorname{log}$, we see that they first enter at the second order through the first term of
	\begin{equation}
		\operatorname{log}\ \operatorname{det}\ \frac{\delta \scrA'}{\delta \scrA}\Bigr|_{\mathcal{O}(g^2)}\=\tr\ \frac{\delta \scrA'}{\delta \scrA}\Bigr|_{\mathcal{O}(g^2)}-\sfrac12 \tr\ \frac{\delta \scrA'}{\delta \scrA}\Bigr|_{\mathcal{O}(g)}\frac{\delta \scrA'}{\delta \scrA}\Bigr|_{\mathcal{O}(g)}\ .
	\end{equation}
	More specifically, we have
	\begin{equation}\label{eq:tr}
		\begin{aligned}
			\tr\ \frac{\delta \scrA'}{\delta \scrA}&\=\int\; \drm^4 x\; \drm^4 y\ \delta^{(4)}(x-y)\delta^{ab}\delta\indices{^\Sigma_\Delta}\frac{\delta \scrA'^a_\Sigma(x)}{\delta \scrA^b_\Delta(y)}\\ &\=\int\; \drm^4 x\; \drm^4 y\ \delta^{(4)}(x-y)\delta^{ab}\Bigl\{\frac{\delta A'^a_\mu(x)}{\delta A^b_\nu(y)}\delta\indices{^\mu_\nu}+\frac{\delta \varphi'^a_I(x)}{\delta \varphi_J^b(y)}\delta_{IJ}+\frac{\delta \varphi'^a_{I+3}(x)}{\delta \varphi_{J+3}^b(y)}\delta_{IJ}\Bigr\}\ .
		\end{aligned}
	\end{equation}
	We need to show that the contributions from the blue terms in \eqref{eq:tr} vanish. We again refer to the three contributions as $\circled{1}$, $\circled{2}$ and $\circled{3}$ respectively and start with the case $A=4$. It is easy to see that
	$\circled{1}=0$,
	since every term contains $\epsilon_{\mu\lambda\rho\sigma}\eta^{\mu\sigma}=0$. For $\circled{2}$, we note that the first line of the blue terms in \eqref{eq:NM_dr2_4} contains no fields $\varphi_{I}$, i.e.~drops out when varying w.r.t.~$\varphi_{I}$. The second term in the last line gives no contribution since $f^{aac}=0$, whereas the remaining two contributions cancel each other. In an analogous fashion, one finds $\circled{3}=0$, so that in total, the blue terms leave the determinant matching condition \eqref{eq:det_mat} invariant. The $A=K$ case works in almost the same way by carrying around the sign factors and additionally taking care of the last line in \eqref{eq:NM_dr2_K}.

\end{document}